\documentclass[12pt]{iopart}

\usepackage{color, soul}
\usepackage{iopams}
\usepackage{graphicx}
\usepackage{cite}
\usepackage[dvipdfm,bookmarks=true,colorlinks=true,linkcolor=blue,urlcolor=blue,
citecolor=blue]{hyperref}

\begin{document}
\title{Charge dynamics of the antiferromagnetically ordered Mott insulator}
\author{Xing-Jie Han,$^1$ Yu Liu,$^2$$^,$$^3$ Zhi-Yuan Liu,$^4$ Xin Li,$^1$ 
Jing Chen,$^1$ Hai-Jun Liao,$^1$ Zhi-Yuan Xie,$^5$ B. Normand$^5$ and Tao 
Xiang$^1$$^,$$^6$}
\address{$^1$Institute of Physics, Chinese Academy of Sciences, Beijing 
100190, People's Republic of China}
\address{$^2$LCP, Institute of Applied Physics and Computational Mathematics,
Beijing 100088, People's Republic of China}
\address{$^3$Software Center for High Performance Numerical Simulation,
Chinese Academy of Engineering Physics, Beijing 100088, People's Republic of 
China}
\address{$^4$Institute of Theoretical Physics, Chinese Academy of Sciences, 
Beijing 100190, People's Republic of China}
\address{$^5$Department of Physics, Renmin University of China, Beijing
100872, People's Republic of China}
\address{$^6$Collaborative Innovation Center of Quantum Matter, Beijing
100190, People's Republic of China}

\ead{txiang@iphy.ac.cn}
\begin{abstract}
We introduce a slave-fermion formulation in which to study the charge dynamics
of the half-filled Hubbard model on the square lattice. In this description,
the charge degrees of freedom are represented by fermionic holons and doublons
and the Mott-insulating characteristics of the ground state are the consequence
of holon-doublon bound-state formation. The bosonic spin degrees of freedom
are described by the antiferromagnetic Heisenberg model, yielding long-ranged
(N\'eel) magnetic order at zero temperature. Within this framework and in the
self-consistent Born approximation, we perform systematic calculations of
the average double occupancy, the electronic density of states, the spectral
function and the optical conductivity. Qualitatively, our method reproduces
the lower and upper Hubbard bands, the spectral-weight transfer into a
coherent quasiparticle band at their lower edges and the renormalisation
of the Mott gap, which is associated with holon-doublon binding, due to the
interactions of both quasiparticle species with the magnons. The zeros of the 
Green function at the chemical potential give the Luttinger volume, the poles 
of the self-energy reflect the underlying quasiparticle dispersion with a 
spin-renormalised hopping parameter and the optical gap is directly related 
to the Mott gap. Quantitatively, the square-lattice Hubbard model is one of 
the best-characterised problems in correlated condensed matter and many 
numerical calculations, all with different strengths and weaknesses, exist 
with which to benchmark our approach. From the semi-quantitative accuracy of 
our results for all but the weakest interaction strengths, we conclude that 
a self-consistent treatment of the spin fluctuation effects on the charge 
degrees of freedom captures all the essential physics of the antiferromagnetic 
Mott-Hubbard insulator. We remark in addition that an analytical approximation 
with these properties serves a vital function in developing a full 
understanding of the fundamental physics of the Mott state, both in the 
antiferromagnetic insulator and at finite temperatures and dopings.
\end{abstract}



\submitto{\NJP}
\maketitle

\section{Introduction}
\label{Sec1}
The mechanism underlying the Mott metal-insulator transition \cite{Imada-1998}
stands as a fundamental theoretical challenge in condensed matter physics. In
1937, de Boer and Verwey \cite{Boer-1937} reported that a class of
transition-metal oxides with partially filled bands, specifically NiO and
MnO, are semiconductors or insulators in direct contradiction to predictions
by conventional band theory. This motivated Mott and Peierls \cite{Mott-1937}
to point out the importance of the electrostatic interaction between the
electrons, and Mott later introduced the concept of the metal-insulator
transition that bears his name \cite{Mott-1949,Mott-1956} to describe
insulating behaviour arising as a result of strong electron-electron
correlations. The discovery of high-$T_c$ superconductivity in a class
of doped antiferromagnetic Mott insulators \cite{Bednorz-1986} revived an
enormous and lasting interest in understanding the Mott phase and the
associated metal-insulator transition.

The Hubbard model \cite{Hubbard-1963} is the minimal model describing the
competition between the kinetic energy of the electrons and their on-site
Coulomb interaction. It captures many characteristic features of strongly
correlated systems and thus serves as a paradigm for numerous phenomena in
condensed matter physics. It is believed that the Hubbard model contains
all the basic physics of the Mott metal-insulator transition and, in some
quarters, that it may reveal the mechanism of high-$T_c$ superconductivity.
However, despite the simplicity of the Hubbard model, exact results can be
obtained only from the Bethe Ansatz \cite{Lieb-1968} in one dimension and
from Dynamical Mean-Field Theory (DMFT) \cite{Metzner-1989,Georges-1996}
in infinite dimensions.

There exist many proposals for the primary mechanism 
driving the Mott transition. Hubbard's equation-of-motion methods
\cite{Hubbard-1963,Hubbard-19642,Hubbard-19643,Hubbard-19654,Hubbard-19675}
attribute the charge gap to the formation of the incoherent lower and upper
Hubbard bands. These provided the first example for a metal-insulator
transition in which the insulating behaviour is not accompanied by the
onset of magnetic order. Brinkman and Rice \cite{Brinkman-19702} applied
the Gutzwiller variational method \cite{Gutzwiller-1965} to treat the
metal-insulator transition out of the Fermi-liquid metallic phase, and
ascribed the transition to the vanishing of the quasiparticle residue,
$Z$, and the divergence of the quasiparticle effective mass, $m^{\ast}$. The
Hubbard approximation captures the incoherent part of the physics while the
Brinkman-Rice approximation captures the coherent part. However, neither
approximation takes the effect of spin fluctuations into account.

For the half-filled single-band Hubbard model on the square lattice, quantum
Monte Carlo simulations \cite{Hirsh-1985,White-1989} have shown that the
ground state is an antiferromagnetic insulator, although by the Mermin-Wagner
theorem its N\'eel temperature is zero. In the weak-coupling limit,
Fermi-surface nesting and the proximity to a van Hove singularity in the
density of states act to induce a spin-density-wave state and thus to produce
a gap \cite{Hirsh-1985}. An asymptotically exact weak-coupling solution for
the Hubbard model was given in reference \cite{Raghu-2010}. In the 
strong-coupling regime, it is the large on-site Coulomb repulsion energy, 
$U$, for double site occupancy that suppresses electron mobility and 
determines the Mott gap.

For the intermediate-coupling regime, where no well-controlled theoretical
solution exists, many numerical methods have been applied to the
two-dimensional (2D) Hubbard model, including exact diagonalisation
\cite{Dagotto-1992,Leung-1992,Feng-1992,Dagotto-1994,Eder-2011}, quantum
Monte Carlo \cite{White-1991,Bulut-1994,Preuss-1995,Hanke-2000,Varney-2009,
Baeriswyl-2009,Yanagisawa-2013,Vitali-2016},
cluster perturbation theory \cite{Senechal-2000,Senechal-2002,Senechal-2004,
Kohno-2012,Kohno-2014,Wang-2015}, the variational cluster approximation
\cite{Potthoff-2003,Dahnken-2004,Tremblay-2005,Hanke-2007,Schafer-2015}
and cluster DMFT \cite{Kotliar-2001,Stanescu-2006,Park-2008,Gull-2011,
Sentef-2011,Sordi-2012}. A detailed review, including further results
from density-matrix renormalisation-group (DMRG) calculations, may be
found in reference \cite{LeBlanc-2015}. However, all of these methods suffer
from different intrinsic limitations. Cluster perturbation theory provides
an approximate lattice Green function for a continuous wave-vector space
but is not self-consistent and cannot describe broken-symmetry states,
which are known to be present for the half-filled square lattice. The
variational cluster approximation can be viewed as an extension of cluster
perturbation theory, which allows for broken symmetries by introducing Weiss
fields, but remains limited by the cluster size. In cluster DMFT, the quantum
impurity model can be solved by quantum Monte Carlo or exact diagonalisation.
The former operates at finite temperature and imaginary time, requiring
extrapolation to recover zero-temperature information and analytic
continuation methods to obtain real-frequency results, neither of which
is well controlled; further, the ubiquitous fermion sign problem affecting
quantum Monte Carlo methods becomes severe when the system is doped. The
latter is implemented at zero temperature and gives direct real-frequency
dynamical information, but can access only small cluster sizes. DMRG is
inherently 1D in nature and can be applied only on a narrow cylinder; the 
ongoing development of higher-dimensional analogues based on tensor-network 
states has progressed to the point where an infinite projected entangled-pair 
state (iPEPS) method has been used very recently to obtain very competitive  
ground-state energies \cite{Corboz-2016}. Anderson \cite{Anderson-1997} has 
argued that the half-filled 2D Hubbard model is fundamentally nonperturbative 
in nature, in the same way as the 1D case, with a Mott gap present for all 
$U > 0$ and robust against temperature. Thus despite all of the theoretical 
and numerical progress made to date, the nature of the Mott gap at half-filling 
and the properties of the 2D Hubbard model remain as challenging open questions.

In the strong-coupling limit, below half-filling the dominant on-site Coulomb
repulsion implies the absolute exclusion of doubly occupied sites. In this
case, the Hubbard model can be mapped to the $t$--$J$ model at the level of
second-order perturbation theory \cite{Auerbach-1994}. At half-filling, no
empty sites remain and this model reduces to the antiferromagnetic Heisenberg
model with only spin-fluctuation degrees of freedom. As $U$ decreases, charge
fluctuations play an increasingly important role in the Hubbard model
\cite{Castellani-1979,Kaplan-1982}, and in the half-filled case the
elementary charge excitations are holons (empty sites) and doublons
(doubly occupied sites), in equal numbers. However, even at weak coupling,
insulating behavior remains guaranteed if the holons and doublons have a
tendency to form bound states, which results in the presence of a charge gap.
Variational Monte Carlo results \cite{Capello-2005,Capello-2006,Yokoyama-2006,
Miyagawa-2011,Miya-2011} have shown that a variational wave function including
holon-doublon binding effects can lower the ground-state energy and that the
Mott transition can be characterised as an unbinding transition of holons and
doublons. Several theoretical proposals for the mechanism of Mott physics
contain holon-doublon binding as an important element, including the ``hidden
charge-2e boson'' mechanism \cite{Leigh-2007,Leigh-2009,Phillips-2010}, the
reconstruction of poles and zeros of the Green function
\cite{Dzyaloshinskii-2003,Sakai-2009,Sakai-2010}, composite fermion
theory \cite{Yamaji-2011} and the Kotliar-Ruckenstein slave-boson theory
\cite{Zhou-2014}. The zeros of the Green function at the chemical potential
in momentum space can be taken to define the ``Luttinger surface,'' which is
closely connected to the non-interacting Fermi surface \cite{Stanescu-2007}.

The motion of a single hole in an antiferromagnetic background has
been studied extensively within the self-consistent Born approximation (SCBA)
\cite{Rink-1988,Kane-1989,Marsiglio-1991,Martinez-1991,Zaanen-1995,Xiang-1996},
where the neglect of Feynman diagrams with crossing propagators is equivalent
to neglecting the distortion of the spin background caused by the presence of
the hole. This formulation is similar to the retraceable path (Brinkman-Rice)
approximation \cite{Brinkman-1970} and the resulting single-hole spectral
function is composed of two components, a sharp peak corresponding to coherent
quasiparticle motion and an incoherent background. The coherent peak arises
from the coupling between the hole and the spin excitations. Recent
experiments have shown that the SCBA yields excellent agreement with
resonant inelastic X-ray scattering measurements performed on the quasi-2D
spin-$1/2$ antiferromagnet Sr$_{2}$IrO$_{4}$ \cite{Kim-2014}.

To make a meaningful contribution to such a complex and deeply studied
problem, here we aim to provide an analytical framework which captures the
essential features of the charge dynamics of the single-band Hubbard model.
To introduce this framework, we restrict our considerations to the
square-lattice model with only nearest-neighbour coupling, to half-filling,
and to zero temperature. While it is unrealistic to expect to find much new
physics in this most generic situation, our goals are to demonstrate the
qualitative power of a suitably chosen mean-field description, to establish
the semi-quantitative accuracy of our results by benchmarking against the
plethora of available numerical studies, and to lay a foundation for
development in the more experimentally relevant directions of finite
temperatures, extended bandstructures and finite dopings.

%\cite{Hirsh-1985,White-1989,Dagotto-1992,Leung-1992,Feng-1992,Dagotto-1994,
%Eder-2011,White-1991,Bulut-1994,Preuss-1995,Hanke-2000,Varney-2009,
%Baeriswyl-2009,Yanagisawa-2013,Senechal-2000,Senechal-2002,Senechal-2004,
%Kohno-2012,Kohno-2014,Wang-2015,Potthoff-2003,Dahnken-2004,Tremblay-2005,
%Hanke-2007,Schafer-2015,Kotliar-2001,Stanescu-2006,Park-2008,Gull-2011,
%Sentef-2011,Sordi-2012,LeBlanc-2015} 

An accurate description of the Mott insulator is based on the strong-coupling
limit, where its properties are robust. As we discuss in more detail below,
this leads to a formalism where the charge degrees of freedom are represented
by fermionic holons and doublons and the Mott-insulating state involves the
formation of holon-doublon bound states. The spin degrees of freedom, 
represented by bosonic magnons, order magnetically at zero temperature for 
any finite interaction strength, but their quantum fluctuations act to 
renormalise the charge sector. Thus the task at hand is to consider the 
antiferromagnetically ordered Mott insulator and to describe accurately 
both the holon-doublon binding process and the spin renormalisation of the 
charge dynamics. 

For specificity, we declare here that we take the on-site interaction to be 
the origin of Mott physics for all dimensionalities, temperatures or dopings, 
and antiferromagnetic fluctuations to be one consequence. It is true that this 
assumption remains unproven for the square lattice (2D) with nearest-neighbour 
hopping at half-filling: in this somewhat pathological case, the perfect 
Fermi-surface nesting means that a gap is opened in the charge sector at any 
finite interaction strength, and in a weak-coupling picture this can be 
interpreted as a process driven by the onset of antiferromagnetic order, 
i.e. not by the charge sector but by the spin sector. This result has led 
to significant confusion over whether an antiferromagnetic insulator can 
exist independently of a Mott insulator. We use the fact that there is no 
transition at any finite interaction strength in the model at hand to deduce 
that the two possible states are different manifestations of the same physics 
and are connected by a crossover. For practical purposes, here we take the 
extensive numerical calculations on the half-filled Hubbard model to indicate 
that the ground state is a magnetically ordered Mott insulator for all 
intermediate (and experimentally relevant) interaction strengths. The Mott 
(charge excitation) gap in our framework is a consequence of holon-doublon 
binding and its presence ensures a finite spectral (single-particle excitation) 
gap, while the spin excitations are gapless. The same holon-doublon binding 
mechanism is equally applicable to the Hubbard model at finite temperature 
or doping, where the spin sector is present only as short-range fluctuations.

Our approach is based on a slave-particle formalism \cite{Barnes-1976,
Barnes-1977,Zou-1988,Yoshioka-1989,Xiang-2009} in which electron operators
are expressed as a combination of ``slave'' fermionic and bosonic operators
preserving the net fermionic statistics, and spin-charge separation is
assumed. A degree of arbitrariness exists in ascribing the fermionic
statistics to the spin (known as the slave-boson approach) or to the charge
degrees of freedom (slave-fermion decomposition). Keeping the importance of
spin fluctuations at the forefront of our considerations, we assume that the
ground state has N\'eel order, which is known in the large-$U$ limit, and
that the spin degrees of freedom are described by the antiferromagnetic
Heisenberg model. The ground state of this model on the square lattice
for $S = 1/2$ is better described by the Schwinger boson (slave-fermion)
formulation \cite{Yoshioka-1989,Manousakis-1991}. Of equal importance, for
an investigation of the holon-doublon binding mechanism it is physically
much more intuitive to ascribe the fermionic statistics to the charge
degrees of freedom, in analogy with the electron binding mechanism of
Bardeen-Cooper-Schrieffer (BCS) superconductivity.

In this slave-fermion framework, antiferromagnetic long-range order
corresponds to the condensation of one of the slave bosons on each
sublattice in the ground state. On this basis, we treat the spin dynamics
within the linear spin-wave approximation \cite{Auerbach-1994}, where the
elementary excitations are magnons. Because the motion of holons and doublons
distorts the antiferromagnetic background even at half-filling, a consistent
account of spin-fluctuation effects is of key importance in describing the
charge dynamics. We treat the interactions among holons, doublons and
magnons within the SCBA to calculate important physical quantities including
the double occupancy, the spectral function, the electronic density of states,
the quasiparticle Green function and the optical conductivity. Our results
show a non-zero double occupancy for any finite $U$, that the Mott-insulating
state results from holon-doublon binding, that spin fluctuations modify the 
size of the Mott gap and that this gap can be probed accurately by measurements 
of the optical conductivity.

This paper is organised as follows. In section~\ref{Sec2} we introduce
formally the model and the methods we use to perform our calculations.
In section~\ref{Sec3} we compute the doublon density for all intermediate 
values of $U$ and in section~\ref{Sec4} we present the spectral function to 
discuss the coherent and incoherent components of the charge response, the 
density of states and the Mott gap. In section~\ref{Sec5} we calculate the 
electron Green function and associated Luttinger surface, and deduce the 
effective quasiparticle bandstructure. Section \ref{Sec6} contains our 
results for and conclusions from the optical conductivity and a summary 
is provided in section~\ref{Sec7}.

\section{Model and Method}
\label{Sec2}

The single-band Hubbard model is defined by the Hamiltonian
\begin{equation}
H = - t \sum_{\langle i,j \rangle \sigma} c_{i\sigma}^{\dagger} c_{j\sigma} + U \sum_{i}
(n_{i\uparrow} - 1/2)(n_{i\downarrow} - 1/2), \label{Hamil}%
\end{equation}
where $c_{i\sigma}$ ($c_{i\sigma}^{\dagger}$) denotes the annihilation (creation)
operator of an electron with spin $\sigma$ on site $i$ of the square lattice
and $\langle i,j \rangle$ indicates that we restrict the hopping terms to
nearest-neighbour sites $i$ and $j$ only. We set the hopping parameter as
$t = 1$ to establish the energy units of our calculations and $U$ represents
the on-site Coulomb repulsion.

In the slave-fermion formalism, the electron operator is written as
\begin{equation}
c_{i\sigma} = s_{i\overline{\sigma}}^{\dagger} d_{i} + \sigma e_{i}^{\dagger}s_{i\sigma},
\label{Sla}
\end{equation}
where $d_{i}$ and $e_{i}$ are fermionic operators denoting the charge degrees
of freedom and $s_{i\sigma}$ are bosonic operators describing the spin
degrees of freedom, with $\sigma = 1$ for spin $\uparrow$ and $-1$ for spin
$\downarrow$. The operator $e_{i}^{\dagger}$ creates an empty (unoccupied) site,
a holon, at lattice point $i$, $d_{i}^{\dag}$ creates a doubly occupied site,
a doublon, and $s_{i\sigma}^{\dagger}$ ($s_{i\bar{\sigma}}^{\dagger}$) is the
creation operator for a singly occupied site $i$ with spin $\sigma$
($\bar{\sigma}$). In this formulation, the local Hilbert space is enlarged
and the constraint
\begin{equation}
d_{i}^{\dagger} d_{i} + e_{i}^{\dagger} e_{i} + \sum_{\sigma} s_{i\sigma}^{\dagger}
s_{i\sigma} = 1
\label{Constraint}
\end{equation}
should be satisfied to eliminate unphysical states.

Substituting equation~(\ref{Sla}) into equation~(\ref{Hamil}) gives the form
\begin{eqnarray}
H & = - t \sum_{i,\delta,\sigma} [(d_{i+\delta}^{\dagger} d_{i} - e_{i+\delta}^{\dagger}
e_{i}) s_{i,\sigma}^{\dagger} s_{i+\delta,\sigma} + \rm{H}.\rm{c}.] \nonumber \\
& \quad - t \sum_{i,\delta,\sigma} [(d_{i}^{\dagger} e_{i+\delta}^{\dagger} + e_{i}^{\dagger}
d_{i+\delta}^{\dagger}) \sigma s_{i,\bar{\sigma}} s_{i+\delta,\sigma} + \rm{H}.\rm{c}.] 
\nonumber \\ & \quad + {\textstyle \frac{1}{2}} U \sum_{i} (d_{i}^{\dagger} 
d_{i} + e_{i}^{\dagger} e_{i} - {\textstyle \frac{1}{2}}), 
\label{HTU}
\end{eqnarray}
where $\delta$ denotes the lattice vectors $(a,0)$ and $(0,a)$. Here we
do not include a Lagrange-multiplier term to enforce the local constraint at
a global level (introducing a chemical potential), but instead implement
equation~(\ref{Constraint}) as a self-consistent condition in our treatment. 
This procedure has the same effect, in that the constraint is satisfied only 
on average, which is a primary shortcoming of all analytical approaches to
locally constrained problems unless gauge fluctuations can be included.
Otherwise, the efficacy of this type of approximation is difficult to assess
by any means other than comparing the results it yields with numerical
calculations where the constraint can be enforced exactly.

From the first line of equation~(\ref{HTU}), holons and doublons can
hop between nearest-neighbour sites with accompanying creation and
annihilation of singly-occupied states. This term serves as the starting
point for applying the SCBA, which provides a proper treatment of the
coupling between the charge degrees of freedom and the spin fluctuations
\cite{Rink-1988,Kane-1989,Marsiglio-1991,Martinez-1991,Zaanen-1995,Xiang-1996}.
This procedure requires the assumption of a N\'eel-ordered ground state, which
is equivalent to a Bose condensation of the $s_{i\sigma}^{\dagger}$ operators and
is discussed in detail below. The second line of equation~(\ref{HTU}) shows 
that the kinetic term of the Hubbard model contains a pairing interaction 
between holons and doublons in the slave-fermion representation. As noted in 
section~\ref{Sec1}, the analogy with BCS theory motivates both the attribution 
of fermionic statistics to the charge degrees of freedom and the formation of 
bosonic bound pairs of fermionic holons and doublons as the process underlying 
the opening of a charge gap and contributing to the charge dynamics of the 
Mott insulator. In the last line of equation~(\ref{HTU}), the Coulomb 
repulsion, $U$, appears as a mass term for holons and doublons, which gain 
a dispersive nature both through their pairing and through their interactions 
with the magnetic background, as described by the SCBA.

As appropriate for a method based on the large-$U$ limit, we assume that the
ground state for the spin degrees of freedom is the N\'eel antiferromagnet,
also for finite $U$, and that their fluctuations are described by the
antiferromagnetic Heisenberg Hamiltonian,
\begin{equation}
H_{S} = J \sum_{\langle i,j \rangle } \mathbf{S}_{i} \cdot \mathbf{S}_{j},
\end{equation}
with the coupling constant taken for simplicity as $J = 4t^{2}/U$. In the
Schwinger boson representation, the spin operators are given by
\begin{equation}
\mathbf{S}_{i} = \frac{1}{2} \sum_{\alpha\beta} s_{i\alpha}^{\dagger} 
\bsigma_{\alpha\beta} s_{i\beta},
\end{equation}
where $\boldsymbol{\sigma} = (\sigma^{x},\sigma^{y},\sigma^{z})$ denotes
the Pauli matrices. The N\'{e}el antiferromagnetic state corresponds to a
Bose-Einstein condensation of one of the two types of bosonic operator
on each sublattice \cite{Hirsch-1989}, which we describe by the uniform
mean-field assumption
\begin{eqnarray}
s_{i\uparrow}^{\dagger}, s_{i\uparrow} & \longrightarrow \langle s_{i\uparrow}
\rangle = \langle s_{i\uparrow}^{\dagger} \rangle = b_{0}, \qquad
i \in A,\nonumber \\
s_{j\downarrow}^{\dagger}, s_{j\downarrow} & \longrightarrow \langle s_{j\downarrow}
\rangle = \langle s_{j\downarrow}^{\dagger} \rangle = b_{0}, \qquad
j \in B,
\label{Condensation}
\end{eqnarray}
for the two sublattices $A$ and $B$. The constraint (\ref{Constraint}) is 
then recast as
\begin{equation}
b_{0}^{2} = \left\{
\begin{array}[c]{c}
1 - \langle d_{i}^{\dagger} d_{i} \rangle - \langle e_{i}^{\dagger} e_{i} \rangle
 - \langle s_{i\downarrow}^{\dagger} s_{i\downarrow} \rangle, \\
1 - \langle d_{i}^{\dagger} d_{i} \rangle - \langle e_{i}^{\dagger} e_{i} \rangle
 - \langle s_{j\uparrow}^{\dagger} s_{j\uparrow}\rangle,
\end{array} \right.
\begin{array} [c]{c}
 i \in A, \\  j \in B.
\end{array}
\label{Consist}
\end{equation}
Because there is only one Schwinger boson on each sublattice, henceforth we
simplify the notation $s_{i\sigma}$ to $s_{i}$. The revised form (\ref{Consist})
of the constraint is the self-consistency condition in our calculation. We
include the spin excitations of the antiferromagnetic Heisenberg model using
the straightforward linear spin-wave approximation; although this method is
only valid strictly for large spin and high-dimensional lattice geometries,
it has been shown \cite{Manousakis-1991} that physical quantities, including
the ground-state energy and sublattice magnetisation, obtained for the
$S = 1/2$ square-lattice model within this approximation are qualitatively
correct and readily renormalised to the values given by numerical
calculations. A more detailed treatment of the Heisenberg model is
provided in~\ref{Heisenberg}.

\begin{figure}[t]
\centering\includegraphics[width=0.6\textwidth]{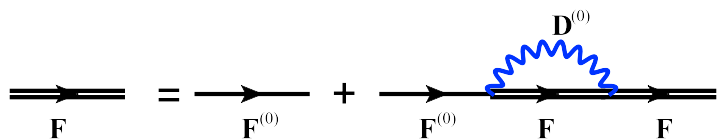}
\caption{Feynman diagrams for the SCBA; fermion and magnon propagators are
represented respectively by straight and wavy lines.}
\label{Feynman}
\end{figure}

Within this approximation, equation~(\ref{HTU}) can be expressed as
\begin{equation}
H = \sum_{\mathbf{k}} \psi_{\mathbf{k}}^{\dagger} \tilde{\varepsilon}_{\mathbf{k}}
\psi_{\mathbf{k}} + \sum_{\mathbf{k},\mathbf{q}} \psi_{\mathbf{k}}^{\dagger} M
(\mathbf{k},\mathbf{q}) \psi_{\mathbf{k}-\mathbf{q}},
\label{SCBAHubbard}
\end{equation}
where
$\psi_{\mathbf{k}}^{\dagger} = ( \!\! \begin{array}[c]{cc} 
d_{-\mathbf{k}+\mathbf{Q}}^{\dagger} & e_{\mathbf{k}} \end{array}  \! )$ is the Nambu 
spinor, which contains the charge degrees of freedom. The explicit forms of 
$\tilde{\varepsilon}_{\mathbf{k}}$ and $M (\mathbf{k},\mathbf{q})$ may be found 
in~\ref{Hubbard} along with full details of the SCBA for the Hubbard model 
in this form. The first term of equation~(\ref{SCBAHubbard}) describes the 
unperturbed charge dynamics, with holon-doublon binding appearing in the 
off-diagonal part of the matrix. The second term describes the interaction 
between the charge and spin degrees of freedom. We define the full charge 
Matsubara Green function as
\begin{equation}
\mathbf{F} (\mathbf{k},\tau) = - \langle T_{\tau} \psi_{\mathbf{k}} (\tau)
\psi_{\mathbf{k}}^{\dagger} (0) \rangle
\label{emgf}
\end{equation}
and calculate this at the level of the SCBA. This process includes the
coupling between the charge and spin dynamics and is equivalent to the series
of Feynman diagrams shown in figure~\ref{Feynman}, where $\mathbf{F}^{(0)}$ and
$\mathbf{F}$ are respectively the bare and interacting charge Green functions
and $\mathbf{D}$ is the magnon Green function, whose explicit form is given
in~\ref{Heisenberg}. The SCBA has been used to calculate the motion
of a single hole in an antiferromagnetic background in the $t$--$J$ model,
where a consistent account of the mutual effects of charge motion and spin
fluctuations is similarly essential. The results of these studies show
that the coupling between the holon and the spin waves induces a
quasiparticle-type response often labelled a spin polaron
\cite{Rink-1988,Kane-1989,Marsiglio-1991,Martinez-1991,Zaanen-1995,Xiang-1996}.
Although a proper treatment of charge and spin fluctuations can be obtained in
this way, we comment that the approximation does not include vertex corrections.

The self-consistent Dyson equation for the charge Green function is calculated
by standard techniques, which yield
\begin{equation}
\mathbf{F} (\mathbf{k},i\omega_{n}) = \frac{1}{i\omega_{n} -
\tilde{\varepsilon}_{\mathbf{k}} - \mathbf{\Sigma} (\mathbf{k},i\omega_{n})},
\label{egf}
\end{equation}
where the self-energy $\mathbf{\Sigma} (\mathbf{k},i\omega_{n})$ is given
in~\ref{Hubbard}. The retarded charge Green function can be obtained
from the Matsubara Green function (\ref{emgf}) by analytic continuation
through $i\omega_{n} \rightarrow \omega + i\eta$. The effect of the parameter
$\eta$ is to provide a finite width to peaks in the density of states.

\begin{figure}[t]
\centering\includegraphics[width=0.6\textwidth]{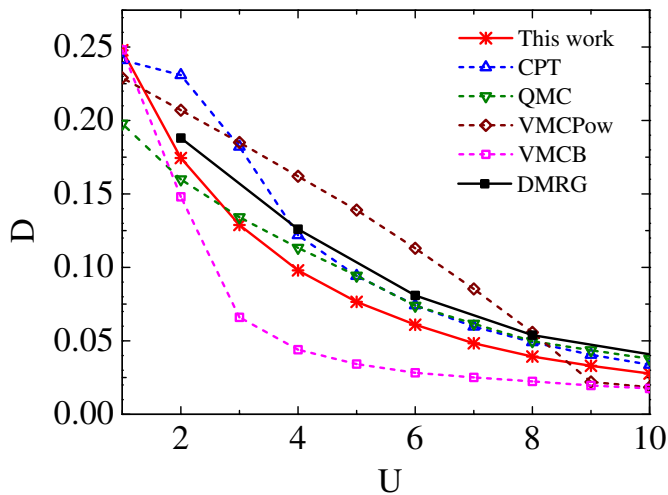}
\caption{Double-occupancy parameter, $D$, shown as a function of on-site 
Coulomb repulsion, $U$. The solid, red line is our result, calculated for a 
system size of $48 \times 48$ and with broadening parameter $\eta = 0.08$. 
For comparison, the dashed lines show analogous results obtained from cluster 
perturbation theory (CPT, blue, up-pointing triangles) \cite{Weng-2014}, 
quantum Monte Carlo (QMC) for a $4 \times 4$ lattice at inverse temperature 
$\beta = 16$ (green, down-pointing triangles) \cite{White-1989}, variational 
Monte Carlo with different trial wave functions [$|\psi_{pow} \rangle$ 
(VMCPow, maroon diamonds) and $|\psi_{B} \rangle$ (VMCB, magenta squares) 
in the notation of reference \cite{Miya-2011}] and DMRG (solid, black line) 
with extrapolation to the thermodynamic limit \cite{LeBlanc-2015}.}
\label{Doublon}
\end{figure}

\section{Double occupancy}
\label{Sec3}

We begin the discussion of our results for the Mott-insulating state of the
square-lattice Hubbard model by considering the averaged double-occupancy
parameter, $D$. For the half-filled system, the average number of doubly
occupied sites is equal to the number of empty ones, and is given by
\begin{equation}
D = \langle n_d \rangle = \frac{1}{N} \sum_{i} \langle d_{i}^{\dagger}
d_{i}\rangle = \frac{1}{N} \sum_{i} \langle n_{i\uparrow} n_{i\downarrow} \rangle.
\end{equation}
$D$ is exactly zero only when the on-site Coulomb repulsion, $U$ (\ref{Hamil}),
is infinite, but charge fluctuations are intrinsic to the Hubbard model for
all finite $U$ values, making $D$ finite. It therefore reflects average
information about the effects of charge fluctuations and as such can be used 
to characterise the Mott transition \cite{Castellani-1979,Yokoyama-2006,
Brinkman-19702,Kotliar-2000,Kaplan-1982,Miyagawa-2011,Miya-2011}.

The dependence of $D$ on $U$ at zero temperature is shown in 
figure~\ref{Doublon}, where our results are compared with calculations by 
quantum Monte Carlo \cite{White-1989}, variational Monte Carlo \cite{Miya-2011},
cluster perturbation theory \cite{Weng-2014} and DMRG \cite{LeBlanc-2015}. At 
the qualitative level, all results lie in the same general range of values and,
with the exception of one VMC approach, show a broadly similar functional
form. Quantitatively, it is immediately clear that the different numerical
results differ from each other quite significantly throughout the regions
of weak and intermediate coupling, which may be taken as a signal of how
difficult the Hubbard problem is in this regime. We stress that finite-size 
extrapolation of our results (not shown) confirms that our $48 \times 48$ 
calculations are fully representative of the infinite system, whereas none 
of the numerical data shown in figure~\ref{Doublon} have been extrapolated 
to the thermodynamic limit other than the DMRG results. However, extrapolated 
results obtained by a range of methods may be found in a very recent review 
\cite{LeBlanc-2015}.

\begin{figure}[t]
\centering\includegraphics[width=0.6\textwidth]{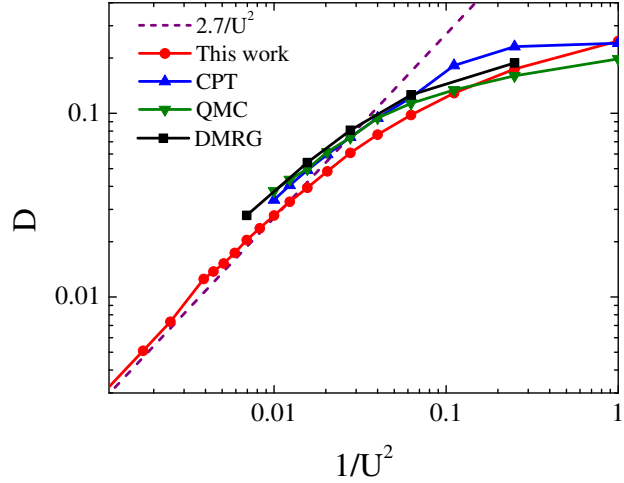}
\caption{Double occupancy parameter, $D$, shown as a function of $1/U^2$ on 
logarithmic axes. Red points show our large-$U$ results, calculated for a 
system size of $48 \times 48$ and with broadening parameter $\eta = 0.08$. 
Results from CPT, QMC and DMRG, shown in the same colour and symbol scheme 
as in figure 2, confirm the trend towards the limiting large-$U$ functional 
form, $D = 2.70/U^2$ (dashed purple line).}
\label{DL}
\end{figure}

Only in the strong-coupling regime do the differences in $D$ values obtained 
from all of these methods become small. Not only is our result in full 
agreement here, but because it is constructed in the strong-coupling limit, 
it can be expected to give the correct asymptotic form of $D$ as $U$ becomes 
large. We have calculated $D$ in the SCBA for a number of large $U$ values up 
to $U = 64$, which we show in figure~\ref{DL}. By considering the doublon 
Green function, $\mathbf{F}^{(0)}_{11} (k,\tau) = \langle d_{-k+Q} (\tau) \, 
d_{-k+Q}^\dag (0) \rangle$, in the limit of large $U$ and extracting the 
zero-temperature doublon occupation from the spectral function according 
to $\langle d_{-k+Q}^\dag d_{-k+Q} \rangle = \int_{-\infty}^0 d\omega A^{(0)} 
(k,\omega)$, one obtains
\begin{equation}
D = \frac{4 t^2 b_0^4}{U^2} \frac{1}{N} \sum_k (\cos k_x + \cos k_y)^2 \simeq
\frac{2.70}{U^2}.
\label{edolu}
\end{equation}
Further details may be found in~\ref{lud}. This asymptotic form is shown in 
figure~\ref{DL}, allowing us to conclude that the double occupation function 
is suppressed at large $U$ according to $D \propto 1/U^2$. The results from a 
number of independent numerical studies, also shown in figure~\ref{DL} are 
fully consistent with $1/U^2$ scaling at large $U$.

At half-filling, the value of $D$ should change monotonically from 0 to
$1/4$, which correspond respectively to the fully localised ($U = \infty$)
and the completely delocalised cases ($U = 0$). As $U$ decreases, charge
fluctuations are enhanced and the on-site Coulomb interaction is screened,
making localisation effects smaller and reducing the Mott gap. The result
is larger $D$ values for smaller $U$, indicating an increasing mixing of
the lower and upper Hubbard bands. Still, it is only at very small values,
$U < 2$, of the on-site repulsion that our results begin to indicate the
breakdown of the approximation, which benchmarks the limits of its
applicability. They do not approach $D = 1/4$ at $U = 0$, which is no
surprise because neither the strong-coupling slave-fermion formalism nor
the single-mode spin-wave approximation to a N\'{e}el antiferromagnetic
ground state as a description of the spin dynamics is appropriate in
this limit.

In general, it is not true that $D$ should increase continuously
with decreasing $U$. In some variational Monte Carlo calculations
\cite{Yokoyama-2006,Miya-2011}, the behaviour of $D$ appears to show
an abrupt change at a critical interaction strength, $U_{c}$, which has
been interpreted as a first-order Mott transition. However, neither our
result nor those obtained from cluster perturbation theory \cite{Weng-2014},
quantum Monte Carlo simulations \cite{White-1989}, or DMRG \cite{LeBlanc-2015}
contain any evidence for a Mott transition at finite $U$. As a consequence of
the perfect nesting, the nearest-neighbour square-lattice Hubbard model is a 
special case where even a weak-coupling treatment gives an insulating state 
for all $U$, i.e.~$U_c = 0$. As noted in section~\ref{Sec1}, this small-$U$ 
result has led to intense debate over the question of whether insulating 
behaviour could be driven by antiferromagnetism rather than by the Coulomb 
interaction, and whether there could be a transition between the two regimes 
at finite $U$. However, the result that $U_c = 0$ in this model means that all 
values of $U$ are continuously connected to the strong-coupling limit, where 
the answers are clear. Indeed, detailed numerical calculations have recently 
been used to argue \cite{Schafer-2015} that the model also has no Mott-Hubbard 
transition at any finite $U$ in the paramagnetic phase at low but finite 
temperatures, meaning that no other effective terms are generated. We conclude 
from our calculations of $D$ that the holon-doublon description yields the 
correct functional form and semi-quantitative accuracy throughout the regime 
of intermediate and strong coupling (specifically, $U > 2$).

\section{Spectral Function}
\label{Sec4}

\subsection{Derivation and Calculation}

We calculate the spectral function of the original electron operators,
$c_{\mathbf{k}\sigma}$, which in the slave-fermion framework are decomposed into
convolutions of the holon operator, $e$, the doublon operator, $d$, and the
Schwinger boson operator, $s$. The electron Green function is defined by
\begin{equation}
G (\mathbf{k},\tau) = - \sum_{\sigma} \langle T_{\tau} c_{\mathbf{k}\sigma} (\tau)
c_{\mathbf{k}\sigma}^{\dagger} (0) \rangle
\end{equation}
and the spectral function
\begin{equation}
A (\mathbf{k},\omega) = - \frac{1}{\pi} \rm{I}\rm{m} G^{R} (\mathbf{k},
\omega + i\eta),
\end{equation}
the imaginary part of the retarded Green function, contains implicitly all
information necessary to describe single-particle excitations. The electron
density of states is obtained from the sum over all wavevectors,
\begin{equation}
\rho (\omega) = \frac{1}{N} \sum_{\mathbf{k}} A (\mathbf{k},\omega).
\end{equation}

\begin{figure}[t]
\centering\includegraphics[width=0.6\textwidth]{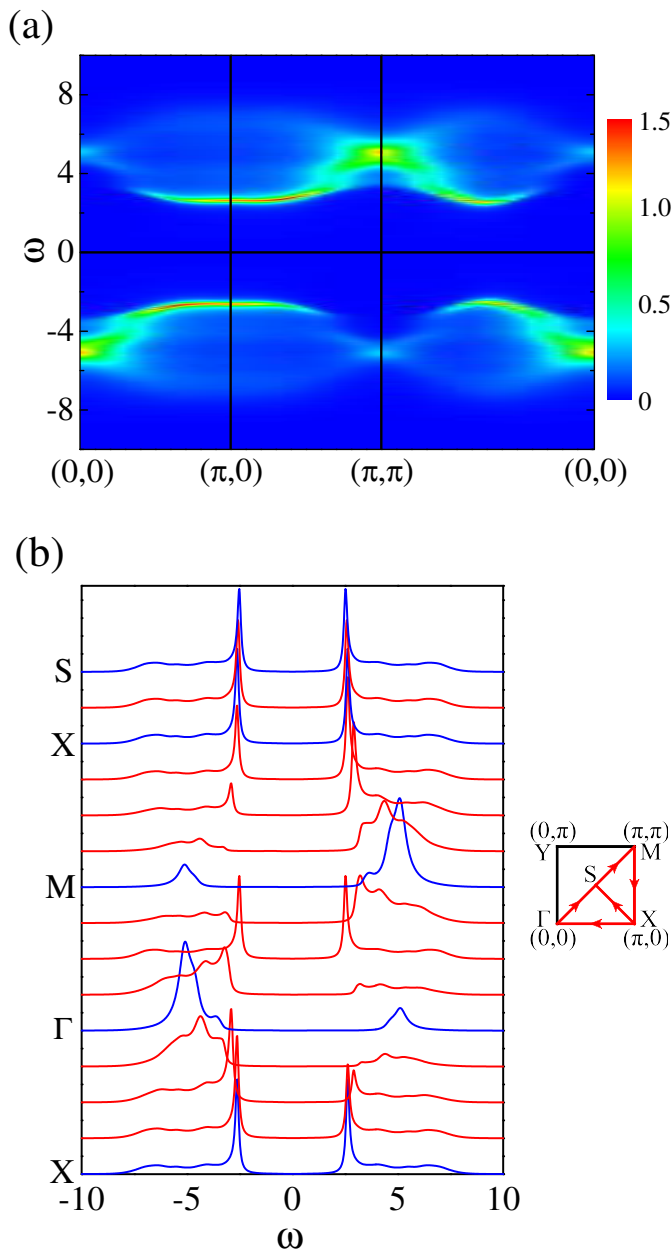}
\caption{Electron spectral function, $A (\mathbf{k},\omega)$, of the Hubbard 
model with $U = 8$, obtained for lattice dimensions $48 \times 48$ and with 
broadening parameter $\eta = 0.08$. (a) Spectral intensity along the symmetry 
directions $\Gamma \rightarrow$ M $\rightarrow$ X $\rightarrow \Gamma$ in the 
first quadrant of the first Brillouin zone. (b) Spectral function $A (\omega)$ 
for wavevectors $\mathbf{k}$ along the high-symmetry directions X $\rightarrow
\Gamma \rightarrow$ M $\rightarrow$ X $\rightarrow$ S.}
\label{Resolved}
\end{figure}

Figure \ref{Resolved} shows the single-particle spectral function,
$A(\mathbf{k},\omega)$, for some specific high-symmetry directions in the
Brillouin zone of the square lattice. These calculations were performed with
$U = 8$ on a $48 \times 48$ lattice and we used a broadening parameter $\eta
 = 0.08$; we stress again that this system size is effectively in the 
thermodynamic limit. The particle-hole symmetry of the spectral function, 
$A(\mathbf{k},\omega) = A(-\mathbf{k+Q},-\omega)$, with $\mathbf{Q} = (\pi,
\pi)$, is preserved. It is clear from the spectral-intensity contour map of
figure~\ref{Resolved}(a) that the spectral weight lies in two separate bands,
the lower and upper Hubbard bands, which are separated by the Mott gap; this
weight appears predominantly in the lower Hubbard band in regions of the
Brillouin zone around $(0,0)$, and is transferred to the upper Hubbard band
as $\mathbf{k}$ moves towards $(\pi,\pi)$.

We begin our comments on the nature of these results by noting that they
are in quantitative agreement with calculations performed by the variational
cluster approximation \cite{Dahnken-2004}, in which the authors included
long-range antiferromagnetic correlations by adding some Weiss fields. That
we obtain the self-consistent interaction effects of the spin fluctuations
on the charge degrees of freedom is one of the key qualitative features of
our approach and will be discussed further below. The shifting of spectral
weight between Hubbard bands signals that the real part of the electron
Green function changes sign between $(0,0)$ and $(\pi,\pi)$, which implies
that the self-energy diverges and the Green function has a zero surface
falling within this region \cite{Dzyaloshinskii-2003}. The importance of
zeros and poles of the Green function has been stressed by many authors
\cite{Eder-2011,Sakai-2009,Sakai-2010,Yamaji-2011,Zhang-2006,
Dzyaloshinskii-2003} in the discussion of Mott physics, particularly in the
context of pseudogap phenomena in the doped system. We defer a discussion of
the zero surface of the electron Green function to section~\ref{Sec5}.

Results for the spectral function, $A(\omega)$, are shown in
figure~\ref{Resolved}(b) for a sequence of wavevectors $\mathbf{k}$
along the high-symmetry directions of the Brillouin zone. The spectral
function is always a multi-peak structure, and the four-peak form reported
by references \cite{Hanke-2000} and \cite{Dahnken-2004} is not very distinctive
in our results. Most of the spectra consist of a peak at low energies,
meaning closer to the chemical potential, accompanied by a broad, weak
high-energy part. The low-energy peak can be interpreted as the coherent
motion of the quasiparticles, which are formed from both holons and doublons
and are renormalise by spin fluctuations. The high-energy part originates in
incoherent charge excitations and forms the remainder of the Hubbard bands.

Returning to the details of figure~\ref{Resolved}(b), we observe that for
$\mathbf{k} = (\pi,0)$ (the X point), the spectral weight is symmetrical
about $\omega = 0$. As $\mathbf{k}$ is moved along X $\rightarrow \Gamma$,
weight is transferred from the positive- to the negative-$\omega$ regime,
i.e.~to lower energies, while along X $\rightarrow$ M the opposite happens;
this behavior is a reflection of particle-hole symmetry. Because ($\pi,0$)
lies on the noninteracting Fermi surface, when $\mathbf{k}$ lies inside this
surface, most of the spectral weight has the properties of a particle, while
outside it has the properties of a hole. An analogous spectral-weight
redistribution is evident as $\mathbf{k}$ is moved from M to $\Gamma$, where
the transfer is from positive to negative $\omega$. At $\mathbf{k} = (\pi/2,
\pi/2) \equiv$ S, the distribution is again symmetrical in $\omega$ and the
gap takes the minimum value obtained in our approximation. We note that all
of these features are very similar to the results of exact diagonalisation
\cite{Dagotto-1992,Leung-1992,Feng-1992,Dagotto-1994} and quantum Monte 
Carlo calculations \cite{White-1991,Bulut-1994,Preuss-1995,Hanke-2000} of 
the Hubbard model at half-filling on the square lattice.

Clearly the leading momentum-dependence of the spectral function is the
consequence of the underlying non-interacting electronic bands. Around X
$\equiv (\pi,0)$, where this dispersion is rather flat
[figure~\ref{Resolved}(a)], the high density of states and strong
interactions are believed to be the origin of the pseudogap and Fermi-arc
phenomena found in doped Hubbard models \cite{Kohno-2012,Kohno-2014}. To
gauge how much of the momentum-dependence may be caused by the coupling
between spin and charge degrees of freedom, we return to the question of
the coherent and incoherent response. For the lower Hubbard band, there is
an apparent suppression of spectral weight around $\omega \approx - 3.5t$,
visible near $\Gamma$, which separates the band into two parts. The
``low-energy'' part (meaning closer to $\omega = 0$) is a
quasiparticle-type band, which disperses from $\Gamma$ to S.

The emergence of a coherent quasiparticle band in the dynamical response is
a highly nontrivial consequence of the interactions between the charge and
spin sectors in the Hubbard model. In general, hole motion in a N\'{e}el
ordered state is energetically unfavourable because the movement distorts the
antiferromagnetic background. However, many calculations within the SCBA
\cite{Rink-1988,Kane-1989,Marsiglio-1991,Martinez-1991,Zaanen-1995,
Xiang-1996} have shown that a quasiparticle, named the ``spin polaron''
\cite{Martinez-1991,Zaanen-1995}, forms as a result of holon-magnon coupling.
The bandwidth of the spin polaron is governed by the spin exchange, $J$, and
this feature forms the low-energy part of the lower Hubbard band. For the
high-energy part, hole motion is thought to originate from effective
three-site hopping processes \cite{Wang-2015}, which allow hole propagation
on the same sublattice that does not distort the antiferromagnetic background
and therefore is less affected by spin fluctuations. We postpone a more
detailed discussion of the Hubbard bands to the next section.

\subsection{Density of States}

\begin{figure}[t]
\centering\includegraphics[width=0.6\textwidth]{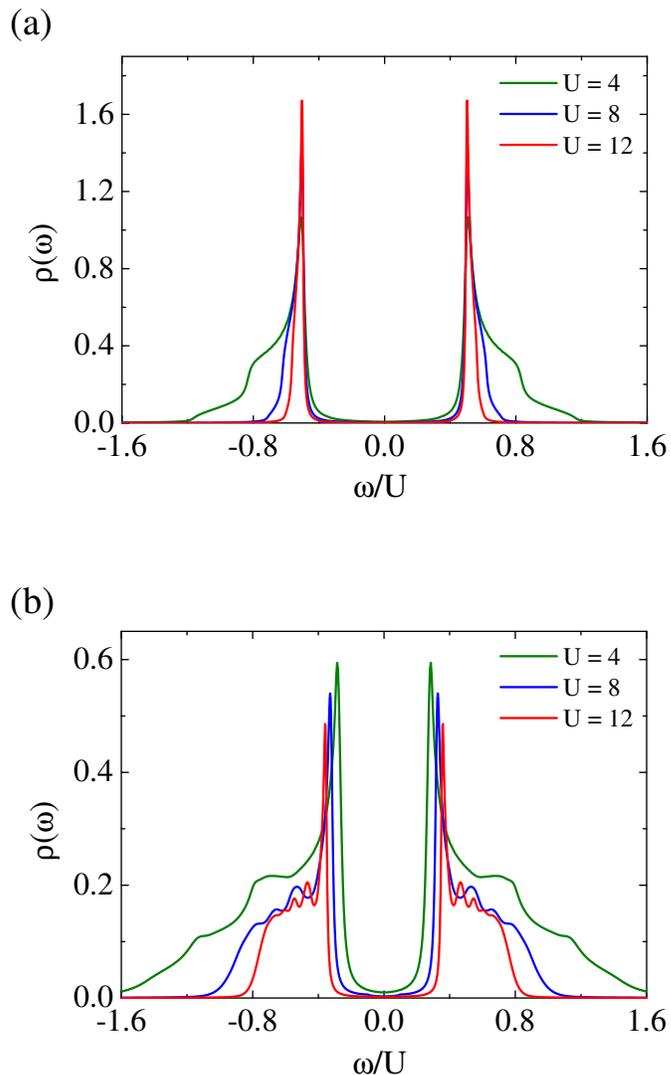}
\caption{Electronic density of states, $\rho (\omega)$, shown for different 
values of $U$; data are normalised to $U$ on the frequency axis. A 
holon-doublon mean-field approximation (a) is compared with the results of 
the SCBA (b).}
\label{Spectral}
\end{figure}

The density of states, $\rho (\omega)$, obtained by integrating the spectral
function is shown in figure~\ref{Spectral}, where we compare the results
calculated in the SCBA with those of a simple holon-doublon mean-field
approximation. The dominant feature in $\rho (\omega)$ is the strong Mott
gap for the values of $U$ illustrated. In the mean-field results, shown in
figure~\ref{Spectral}(a), the effects of spin fluctuations are neglected and
the Mott gap is $U$. However, in our calculations [figure~\ref{Spectral}(b)]
this gap is renormalised downwards, quite significantly at intermediate
values of $U$. This renormalisation is related directly to the other
significant feature of $\rho (\omega)$, the width of the lower and upper
Hubbard bands, which clearly broadens (relative to $U$) as the on-site
interaction decreases and spin fluctuations strengthen. As we will discuss
in section~\ref{Sec5}, the bandwidth reflects the spin-related renormalisation
of the underlying quasiparticle bands and may be treated as an effective
hopping parameter, $t_{\rm eff}$, which is a fraction of the electron hopping
parameter, $t$; by contrast, the Hubbard bands given by the mean-field
approximation have a width given only by the spin interaction, $J = 4t^2/U$,
and therefore appear much narrower in figure~\ref{Spectral}(a).

In the SCBA, only the quasiparticle band remains of width $J$. In general,
the spin polaron is a feature of the retraceable path approximation
\cite{Brinkman-1970} and is well-defined in a mean-field treatment with
high lattice coordination. The energetics of the Mott gap and the Hubbard
bands have been discussed in the DMFT, which is infinite-dimensional, and
the same physics of a narrow quasiparticle band, a renormalised Mott gap
and broader Hubbard bands is found \cite{Sangiovanni-2006}. That our 2D
holon-doublon description reproduces these features with high accuracy
serves both as an indication of the degree to which it captures the key
ingredients of Mott physics and as a means to verify the relevance of
features observed in numerical calculations with different limitations. In
the context of DMFT comparisons, we caution that the small peaks visible
around $\omega/U \approx 0.5$ in figure~\ref{Spectral}(b) are size effects,
which we can demonstrate by smoothing them away if the broadening factor
$\eta$ is set to a larger value.

Although a number of experiments have demonstrated the insulating behaviour
of the parent compounds of the high-$T_{c}$ superconductors, it has proven
to be very difficult to observe both the lower and upper Hubbard bands.
Angle-resolved photoemission spectroscopy \cite{Damascelli-2003} probes the
occupied states while optical spectroscopy \cite{Basov-2005,Basov-2011} gives
information concerning two-particle correlations. Measurements by
scanning tunnelling spectroscopy (STS) probe the local single-particle
density of states for both occupied and unoccupied bands. Only recently
was the full electronic spectrum for the undoped Mott insulator
Ca$_{2}$CuO$_{2}$Cl$_{2}$ measured by STS \cite{Wang-2013}, showing the
Mott-Hubbard gap, and in particular the upper Hubbard band, for the first
time. A comparison with the robust features of our results, namely the spin
polaron, the renormalised Mott gap (section~\ref{Sec43}), and the width of 
the Hubbard bands scaling with $t$, implies that the Hubbard model does in
fact capture the essential properties of the undoped Mott insulator
measured by STS.

\begin{figure}[t]
\centering\includegraphics[width=0.6\textwidth]{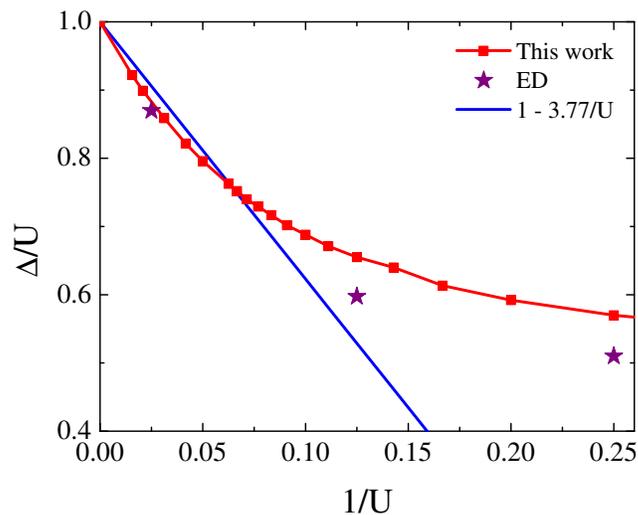}
\caption{Mott gap, $\Delta$, normalised to $U$ and shown as a function 
of $1/U$ to illustrate its asymptotic large-$U$ limit. The purple stars 
indicate results for $U = 4$, 8 and 40 extracted from the exact 
diagonalisation (ED) calculations of reference \cite{Feng-1992}.}
\label{fmglu}
\end{figure}

\subsection{Mott Gap in the Large-$U$ Limit}
\label{Sec43}
We reiterate that the SCBA includes the interaction effects of the holons
and doublons with the magnons, causing significant renormalisation effects
at intermediate values of $U$. Only in the limit of large $U$ do these
effects vanish, causing the Mott gap to tend towards $U$ and the bandwidths
to vanish. We define the Mott gap, $\Delta$, from the separation of the peaks 
in the density of states shown in figure \ref{Spectral}(b) and present the 
results in figure \ref{fmglu}. We comment that, although this definition is 
completely valid in the large-$U$ regime, it may be called into question for 
values of $U/t$ at the far right-hand side of figure \ref{fmglu}: for $U/t
 = 4$ it is clear in figure \ref{Spectral}(b) that there is a significant 
density of states ``inside the Mott gap'' and a detailed account of its
effects could lower the estimate of the effective $\Delta$ \cite{Vitali-2016}.
Spectral information is significantly more difficult to obtain from numerical 
calculations than are ground-state properties, and in figure \ref{fmglu} we 
show only Mott-gap estimates obtained from exact-diagonalisation studies 
\cite{Feng-1992}. These lie very close to our SCBA results at large $U$ and 
then fall with a similar form, but to values slightly smaller than SCBA, as 
$U/t$ is decreased through the intermediate-coupling regime. 

At the left-hand side of figure \ref{fmglu}, similar to our extraction of 
the limiting behaviour of $D$ in section~\ref{Sec3}, we may also investigate 
analytically the approach of the Mott gap to $U$ as $U$ becomes large. As 
shown in~\ref{lug}, by neglecting off-diagonal terms ($\mathbf{F}_{12}$ and 
$\mathbf{F}_{21}$) in the Green function of equation~(\ref{egf}) and using the 
bare holon and doublon Green functions to obtain the spectral function, and 
hence the self-energies, one may construct the full Green function at the 
level only of the first step in the Born approximation. The Mott gap is 
simply the minimum of the holon and doublon gaps, which are identical and, 
on retaining only terms of order unity, are given by $\Delta_{\mathbf{k}} = 
U - 2 \sqrt{a_{\mathbf{k}}}$, where $a_{\mathbf{k}} = \sum_{\mathbf{q}} g_1^2 
(\mathbf{k},\mathbf{q})$ [equation~(\ref{gfunc})]. Thus one obtains
\begin{eqnarray}
\Delta & = & {\rm min}|_{\mathbf{k}} [\Delta_{\mathbf{k}}] = {\rm min}|_{\mathbf{k}}
\left[ U - \sqrt{ 2 \sum_{\mathbf{q}} g_1^2(\mathbf{k},\mathbf{q})} \right]
\\ & = & {\rm min}|_{\mathbf{k}} \left[ U - \sqrt{\frac{2 t^2 b_0^2 z^2}{N} 
\sum_{\mathbf{q}} (u_{\mathbf{q}} \gamma_{\mathbf{q-k}} + v_{\mathbf{q}} 
\gamma_{\mathbf{k}})^2} \right] \nonumber \; \simeq \; U - 3.77, \nonumber
\label{emglu}
\end{eqnarray}
where the gap minimum occurs at $\mathbf{k} = (\pi/2,\pi/2) =$ S. We show
this function in figure~\ref{fmglu} in the form $\Delta/U = 1 - 3.77/U$. The
agreement is satisfactory, confirming that the limiting form of the approach
is indeed $1 - \Delta/U \propto 1/U$, although the constant of proportionality
could be computed more accurately by retaining more orders in the Born
approximation. We comment that this type of limiting behavior has been 
suggested in some numerical studies \cite{Sangiovanni-2006}. It is also 
worth noting \cite{Vitali-2016} that the Mott gap obtained in 1D from the 
Bethe Ansatz has the large-$U$ limit $\Delta/U = 1 - 4/U$ to first order in 
$1/U$. 

In the high-$U$ (atomic) limit, the spectral function is entirely incoherent.
The mean-field approximation gives the ``unperturbed'' charge Green function,
meaning in the absence of spin fluctuations, and the two Hubbard bands in
figure~\ref{Spectral}(a) are also incoherent. In a self-consistent treatment,
the spin fluctuations cause a spectral-weight transfer from the incoherent
background into a coherent quasiparticle band, allowing both the
(incoherent) lower and upper Hubbard bands and the (coherent) low-energy
quasiparticle band to be reconstructed, as shown in figure~\ref{Resolved}(a).
These results emphasise again that a consistent inclusion of the spin degrees
of freedom is intrinsic to a full understanding of the Mott state.

\begin{figure}[t]
\centering\includegraphics[width=0.6\textwidth]{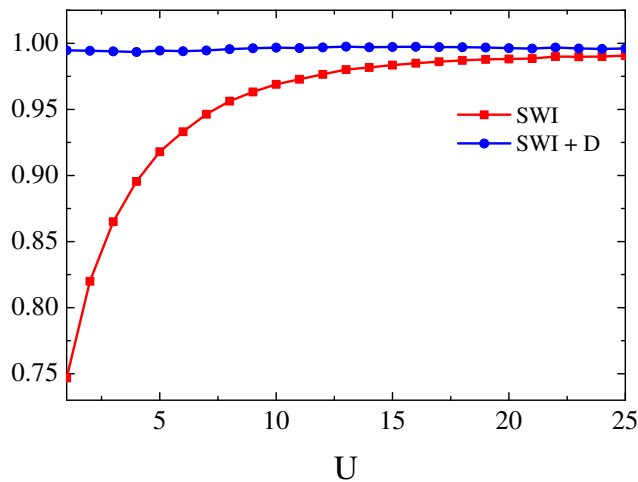}
\caption{Spectral-weight integral [equation~(\ref{eswic})] shown alone 
(red) and corrected by the average double occupancy, $D$ (blue).}
\label{SWI}
\end{figure}

\subsection{Spectral-Weight Integral}

We comment here that, from the properties of the Green function, the density
of states of electrons with both spin orientations should obey the sum rule
\begin{equation}
\int_{-\infty}^{\infty} d \omega \rho (\omega) = 1.
\end{equation}
In our results, because one of the two bosonic operators on each sublattice
undergoes Bose-Einstein condensation [equation~(\ref{Condensation})], the 
sum rule is violated and instead takes the form
\begin{equation}
\int_{-\infty }^{\infty } d \omega \rho (\omega) = 1 - D,
\label{eswic}
\end{equation}
where $D$ is the average double-occupancy parameter calculated in 
section~\ref{Sec3}. In figure~\ref{SWI} we show both the spectral-weight 
integral and its value corrected by the average double occupancy, for a range 
of values of $U$. Because our calculation is performed using a finite energy 
interval, the corrected spectral-weight integral is less than 1, but it is 
clear that the deviations are very small for all intermediate values of $U$. 
In the atomic limit, $D = 0$, the sum rule is satisfied by the spectral weight 
alone, while for finite $U$ the spectral-weight integral violates the sum rule 
by an amount equivalent to $D$ and approaches $1$ as $U$ increases. For an 
intermediate interaction strength such as $U = 8$, we observe that the 
spectral-weight integral is $0.96$, indicating that our strong-coupling 
approximation remains robust, and capable of capturing the primary physics 
of the Hubbard model, in this regime. Only at weak coupling, where the 
spin-wave treatment of the Heisenberg model is not suitable, higher-order 
perturbations are important and charge fluctuations become strong, does 
the SCBA fail to reproduce the complex charge dynamics.

\section{Luttinger Volume}
\label{Sec5}

Landau's Fermi-liquid theory is one of the fundamental theories in
quantum many-body physics. Its most important single qualitative result is
that the Fermi surface still exists, even in the presence of interactions,
in all dimensions higher than 1. For the normal metals that can be
described as Fermi liquids, the shape of the Fermi surface is essential
in determining the thermodynamic and transport properties of the system,
because only electrons close to the Fermi surface can be excited at low
energies. For a given electron density, regardless of the shape of the Fermi
surface, the volume it encloses is invariant even when interactions between
particles are taken into consideration. This result, the Luttinger theorem
\cite{Luttinger-1960,LuttingerW-1960,Oshikawa-2000}, relates the total
electron density to the volume in momentum space where the Green function
is positive,
\begin{equation}
\frac{N}{V} = 2 \int_{G(\mathbf{k},\omega = 0) > 0} \frac{d^{d}k}{(2\pi)^{d}},
\end{equation}
with $d$ the spatial dimension of the system. Although the original proof
by Luttinger was based on perturbation theory, and thus the theorem may in
principle be violated by nonperturbative effects, Oshikawa \cite{Oshikawa-2000}
has provided a nonperturbative proof based on topological arguments. In
essence, the insensitivity of the result to any interactions should be
considered as a quantisation phenomenon \cite{Oshikawa-2000}, and therefore
the Luttinger theorem may be the first example of topological quantisation
discovered for the quantum many-body problem.

In general, the poles of the electron Green function, $G (\mathbf{k},
\omega)$, determine the band dispersion and define the Fermi surface at
the chemical potential, where $\omega = 0$. From the asymptotic behaviour
$G(\mathbf{k},\omega \rightarrow \infty) \propto 1/\omega$, it is obvious
that the limiting Green function is positive for positive $\omega$ and
conversely. For a Fermi liquid, as the frequency approaches the Fermi
surface ($\omega = 0$), by definition $G(\mathbf{k},\omega = 0^{+})
\rightarrow + \infty$ while $G(\mathbf{k},\omega = 0^{-})  \rightarrow
 - \infty$, implying that the Green function must change sign at the Fermi
surface through an infinity in $G(\mathbf{k},0)$ if there are no singularities
in any other regions of $\mathbf{k}$ and $\omega$. The Luttinger theorem then
concerns precisely the volume enclosed by the surfaces where $G(\mathbf{k},0)$
changes sign.

\begin{figure}[t]
\centering\includegraphics[width=0.6\textwidth]{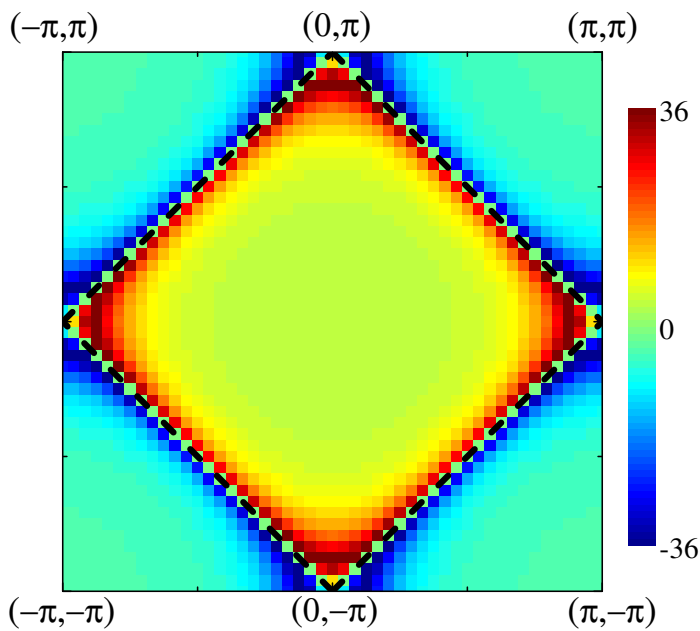}
\caption{Real part of the inverse Green function, $\textrm{Re} [G^{-1} 
(\mathbf{k},\omega = 0)]$, for the half-filled Hubbard model on the square 
lattice, calculated for $U = 8$. $\textrm{Re} [G^{-1} (\mathbf{k},\omega = 0)]$
changes sign from a negative to a positive maximum through a diamond-shaped
boundary. The sign-change gives the Luttinger surface at $\omega = 0$, and
coincides with the locus of solutions of $\cos k_{x} + \cos k_{y} = 0$, marked
by the black, dashed line.}
\label{ZeroSurface}
\end{figure}

However, this change of sign is not always connected with an infinity. It
may also occur through zeros of the Green function \cite{Eder-2011,
Dzyaloshinskii-2003,Sakai-2009,Sakai-2010,Yamaji-2011,Zhang-2006}, and in
fact these are the relevant quantities for insulating systems, where there
is no Fermi surface. By contrast, all systems have a Luttinger surface, and
this may be defined from the zeros. Examples of this type of physics include
the BCS superconductor, where $G(\mathbf{k},0)$ changes sign through the
chemical potential in the absence of a Fermi surface, and also the Mott
insulator.

For systems with a full gap spanning some energy interval
(figures~\ref{Resolved} and \ref{Spectral}), the self-energy inside the
gap must be infinite as otherwise spectral features would appear. For the
Green function, the divergence of the self-energy implies (\ref{egf}) that
$G(\mathbf{k},\omega) = 0$ and thus the poles of the self-energy can be used
to define the Luttinger surface. Here we use the real part of the inverse
Green function to calculate the Luttinger surface, which in terms of
$\textrm{Re} [G^{-1} (\mathbf{k},\omega)]$ corresponds to its points of
divergence, whereas the band dispersion is given by its zeros. We note that
the self-energy can have only one pole, $\xi_{k}$, in an energy regime where
a gap is present; a situation with two successive poles, $\xi_{k,1} < \xi_{k,2}$,
would give $\textrm{Re} [G^{-1}(\mathbf{k},\xi_{k,1} + 0^{+})] \rightarrow
 + \infty$ and $\textrm{Re} [G^{-1}(\mathbf{k},\xi_{k,2} + 0^{-})] \rightarrow
 - \infty$, which would require a zero of $\textrm{Re} [G^{-1}(\mathbf{k},
\omega)]$ in the gap region, contradicting the definition.

In figure~\ref{ZeroSurface} we show the real part of the inverse Green function
in the form of colour contours. The finite broadening factor, $\eta$, in our
calculation converts the divergence into a sharp minimum followed by an
equally sharp maximum as a function of $|\mathbf{k}|$. It is clear that
$\textrm{Re} [G^{-1}(\mathbf{k},\omega = 0)]$ changes sign from negative to
positive through a diamond-shaped boundary, which therefore marks
the zeros of the Green function at $\omega = 0$, i.e.~the Luttinger surface.
This surface is specified precisely by the condition $\cos k_{x} + \cos k_{y}
 = 0$, and therefore we have found that the Luttinger surface is exactly
the Fermi surface of non-interacting electrons. This result is a consequence
of particle-hole symmetry at half-filling. Although the system is completely
gapped in the Mott-insulating phase, the Luttinger theorem remains applicable
and for the half-filled Hubbard model can be interpreted as the integral of
the region within the Luttinger surface \cite{Stanescu-2007}.

\begin{figure}[t]
\centering\includegraphics[width=0.6\textwidth]{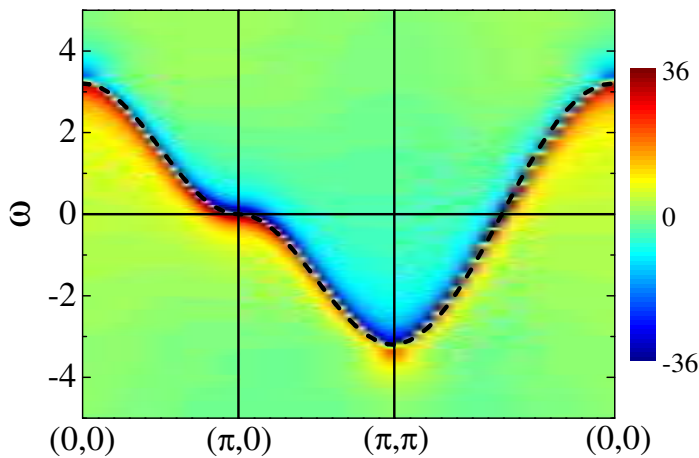}
\caption{Real part of the inverse Green function, $\textrm{Re} [G^{-1} 
(\mathbf{k},\omega)]$, calculated using $U = 8$ for values of $\mathbf{k}$ 
chosen along the high-symmetry directions. The black, dashed curve shows the 
function $1.6t (\cos k_{x} + \cos k_{y})$ (see text).}
\label{ReGreverse}
\end{figure}

To gain more insight into the behaviour of the self-energy, in
figure~\ref{ReGreverse} we show $\textrm{Re} [G^{-1}(\mathbf{k},\omega)]$
calculated with $U = 8$ for the primary high-symmetry directions in momentum
space. The divergence of the inverse Green function can be regarded as a
dispersion relation defined by the self-energy, and it forms a rather clear
quasiparticle band. The black, dashed curve in figure~\ref{ReGreverse} shows
the function $2 t_{\rm eff} (\cos k_{x} + \cos k_{y})$ with $t_{\rm eff} = 0.8 t$,
which obviously provides a reasonable, if not perfect, fit to the zeros of
the calculated Green function. This form, which resembles an inverted
free-electron dispersion with a renormalised hopping parameter, is quite
similar to the results obtained by exact diagonalisation \cite{Eder-2011},
except in that the effective hopping is smaller. 

\begin{figure}[t]
\centering\includegraphics[width=0.6\textwidth]{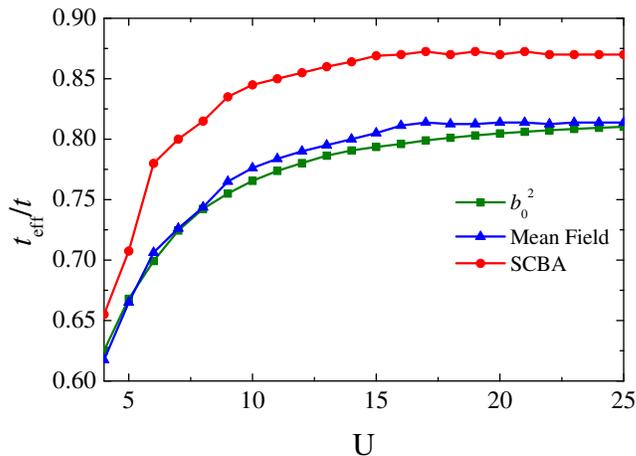}
\caption{Effective hopping parameter, $t_{\rm eff}$, defined by the poles of 
the electron self-energy, $2 t_{\rm eff} (\cos k_{x} + \cos k_{y})$, calculated 
in a mean-field approximation with no spin-fluctuation effects (blue) and in 
the SCBA (red); for comparison we show also the condensation parameter, 
$b_0^2$ (green).}
\label{Hopping}
\end{figure}

To interpret this result, we note first that the inverted nature of the
quasiparticle band signifies its holonic origin. From equation~(\ref{Momentum}),
we expect that the doublon band will have the same form as the holon band
but with a momentum shift of $\mathbf{Q}$. Thus the poles of the electron 
self-energy are given directly by the dispersion relation of the 
quasiparticles, which is in turn connected with the non-interacting 
dispersion, $2 t (\cos k_{x} + \cos k_{y})$, by a renormalisation factor. 
Second, from equation~(\ref{edisp}) one observes that this renormalisation 
of the effective quasiparticle dispersion, $4 t b_{0}^{2} \gamma _{\mathbf{k}}
 = 2 t b_{0}^{2} ( \cos k_{x} + \cos k_{y})$, is related directly to the 
holon-doublon pairing interaction term, which in turn is determined by the 
magnetic ordering (condensation) parameter, $b_{0}$. In the limit of large 
$U$, the Mott gap saturates at $\Delta = U$ (section~\ref{Sec43}) and, as 
discussed in~\ref{lum}, magnetic order becomes robust, with an ordered moment 
of $m_s = 0.321$ and, from equation~(\ref{Consist}), $b_{0}^{2} = 0.821$ in our 
calculations. The effective quasiparticle hopping parameter, $t_{\rm eff}$, is 
shown in figure~\ref{Hopping}, which compares the results extracted from the 
Green function in a simple mean-field approximation, with only holon-doublon 
interactions but without spin-fluctuation effects, to the results of the SCBA. 
Also shown for comparison is the result of a direct determination of $b_0^2$, 
which allows us to deduce that the effective hopping parameter saturates at
$t_{\rm eff} = b_0^2 t = 0.821 t$ as $U \rightarrow \infty$ in the mean-field 
approximation. However, the effect of the additional spin fluctuations 
contained in the SCBA is to allow the effective hopping to be stronger 
than the value constrained by the condensation parameter.

Away from strong coupling, the departure of the quasiparticle mass from
the high-$U$ limit may be taken as a further indication of the effects of
charge and spin fluctuations at intermediate $U$, which includes the value
$U = 8$ shown in figure~\ref{ReGreverse}. The reduction of $t_{\rm eff}$ computed
in the SCBA with decreasing $U$ is clear in figure~\ref{Hopping}, where again
the comparison with the mean-field form and $b_0^2$ allows us to benchmark
the effects of spin fluctuations and of self-consistency at intermediate $U$
within our calculation. The Mott gap is effectively the mass term, $U$, in
equation~(\ref{edisp}) and its downward renormalisation with decreasing $U$ is
also the consequence of holon-doublon interactions, renormalised at
intermediate $U$ by spin fluctuations, as discussed in section~\ref{Sec4} and
shown in figures~\ref{Resolved} and \ref{Spectral}. However, this information
is not visible in the inverse Green function, which therefore can be taken
as a measure only of the spin-fluctuation renormalisation effect within the
SCBA, independently of the Mott-gap scale.

We conclude that the holon-doublon description contains valuable insight
into the static and dynamic properties of the charge degrees of freedom. The
characteristic feature of the Mott insulator is not the Fermi surface but
the Luttinger surface. The Luttinger theorem for the Fermi liquid holds for
the half-filled Hubbard model, despite its insulating nature, with the
sign-change of the Green function at $\omega = 0$ caused by its zeros
instead of its poles. This Luttinger surface appears inside the Mott gap,
which separates the lower and upper Hubbard bands, and it defines a
quasiparticle band whose renormalisation characterises the effects of
spin interactions on the charge sector.

\section{Optical Conductivity}
\label{Sec6}

The final quantity we calculate is the optical conductivity. Optical
conductivity measurements \cite{Basov-2005,Basov-2011}, which reveal the
carrier number, the size of the energy gap, the dynamics of quasiparticle
excitations and their scattering processes, have played an important role
in the study of many classes of correlated electronic materials. For the
insulating parent compounds of the high-$T_{c}$ superconductors, optical
conductivity measurements performed on La$_{2}$CuO$_{4}$ at a temperature
of $300$ K show an excitonic absorption peak at $2$ eV \cite{Uchida-1991}.
This behaviour was ascribed to the strongly correlated but
charge-transfer-dominated character of the undoped CuO$_{2}$ plane
\cite{Zaanen-1985}. When holes are doped into the system, optical
conductivity measurements show a number of features that deviate strongly
from conventional band theory. The most striking phenomenon is the
reconstruction of the electronic spectral weight at low doping, which
involves the transfer of weight from the charge-transfer excitation
regime to a mid-infrared band, centred at approximately $0.5$ eV. This
mid-infrared band is consistent with the appearance of in-gap states
\cite{Weng-2014,Wang-2013} but to date there is no theory for its
microscopic origin. As the doping is increased, a Drude-type response
develops at far-infrared frequencies and decays much more slowly than
band theory would predict. This characteristic spectral-weight transfer
is evidently intrinsic to the Mott insulator and cannot be described by
a theory ignoring electronic correlations. It has been argued
\cite{Phillips-2010,Meinders-1993} that this property of the Mott phase
can be explained within the Hubbard model, but there is as yet no
consensus on the complete dynamics of the weight-transfer phenomenon.

\begin{figure}[t]
\centering\includegraphics[width=0.6\textwidth]{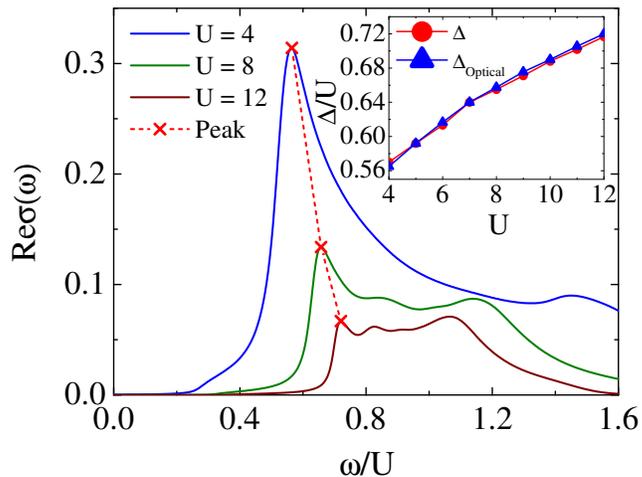}
\caption{Real part of the optical conductivity as a function of frequency,
shown for a number of $U$ values. The results were calculated on a $48 \times
48$ lattice at zero temperature. The red, dashed line connects the initial
peaks on each curve, which determine the optical gap, $\Delta_{\rm Optical}$.
The inset compares $\Delta_{\rm Optical}$ with the Mott gap, $\Delta$,
determined from the peak separation in figure~\protect{\ref{Spectral}}.}
\label{Optical}
\end{figure}

The optical conductivity in the $x$ direction can be calculated from the
imaginary part of the current-current correlation function,
\begin{equation}
\mathrm{Re} [\sigma_{xx} (\omega)] = - \mathrm{Im} \left[\frac{1}{\omega}
\Pi_{xx}^{R} (\omega) \right],
\end{equation}
as discussed in detail in~\ref{Opticalconductivity}. The results
we obtain within the slave-particle framework and the SCBA are shown
in figure~\ref{Optical} for a number of $U$ values spanning the
intermediate-coupling regime. The optical conductivity has a
clear charge-excitation gap, as expected for an insulating system (but
in contrast to the Drude peak predicted by conventional band theory). The
first peak is expected at the ``optical gap,'' $\Delta_{\rm Optical}$, which
is determined by a two-particle process; as $U$ increases, its location
clearly moves to higher energy, even when normalised to $U$, and its peak
height drops. This is consistent with experimental results for the optical
reflectivity spectra \cite{Uchida-1991}, and with the expectation of a larger
quasiparticle gap with increasing $U$ contained in figure~\ref{Spectral}. The
Mott gap is the one-particle charge gap and is determined from the separation
of the leading peaks in the lower and upper Hubbard bands, as shown in the
spectral function [figure~\ref{Resolved}(a)] and the density of states
[figure~\ref{Spectral}(b)]. The inset of figure~\ref{Optical} compares the 
Mott gap with the optical gap obtained in the main panel, and it is clear 
that the two are almost identical for the Mott insulator.

Exact diagonalisation results \cite{DagottoE-1992,Tohyama-2005,Dagotto-1994}
and quantum Monte Carlo calculations \cite{Bulut-1994} also confirm the
insulating character of the undoped Hubbard model. The optical conductivity
given by exact diagonalisation \cite{Tohyama-2005} shows a sharp peak at
the Mott-gap edge, which is attributed to spin-polaron formation in the
photoexcited state, as also revealed by DMFT \cite{Taranto-2012}. Although
these results suggest a separation of the spin-polaron band from the
high-energy feature, we caution that the system sizes used are very small;
in our calculations, the spectral function has a continuous and multipeak
structure for most wavevectors $\mathbf{k}$, with no clear band separation
and thus no discrete peaks in the optical conductivity. We do not attempt a
quantitative comparison of our results with other numerical methods because
the effect of vertex corrections \cite{Lin-2009,Bergeron-2011} cannot be
neglected in the calculation of optical conductivity.

As a consequence of fundamental conservation laws, the optical conductivity
obeys the sum rule \cite{Basov-2005,Basov-2011,Bergeron-2011}
\begin{equation}
\int_{0}^{\infty} d \omega \mathrm{Re} [\sigma_{xx} (\omega)] = - \frac{\pi}{2}
\langle T_{x} \rangle,
\end{equation}
where $\langle T_{x} \rangle$ is the average kinetic energy in the $x$
direction. In our calculations for the half-filled Hubbard model, the
results do not satisfy this sum rule exactly, although the deviation is
within $10\%$ for all intermediate $U$ values. This discrepancy is presumably
caused by the spectral-weight loss discussed in section~\ref{Sec4}, as well 
as by our neglect of vertex corrections. The inclusion of the latter, and of 
higher-order physical processes, may be expected to produce superior results, 
and to be increasingly important for calculations at lower $U$ or at any 
finite temperatures.

\section{Summary}
\label{Sec7}

We have used a holon-doublon slave-fermion representation in the
self-consistent Born approximation to investigate the charge dynamics of
the single-band Hubbard model at half-filling on the square lattice. We show
that holon-doublon binding is intrinsic to the Mott-insulating state. As
expected in a nearest-neighbour hopping model, the system is always insulating
and the monotonically decreasing doublon density shows no critical value,
$U_{c}$, where a metal-insulator transition occurs. The electronic density of
states we compute has a coherent quasiparticle band accompanied by incoherent
lower and upper Hubbard bands. The Mott gap is the excitation gap of the
holons and doublons, and so is determined by their binding energy. We
calculate the Luttinger surface of the Green function at the chemical
potential and demonstrate that, for the Mott insulator, the Luttinger
theorem can be interpreted as the integral of the region enclosed by this
Luttinger surface rather than a Fermi surface. We show further that the
poles of the self-energy define an effective quasiparticle dispersion
similar to that of free electrons, but with a renormalised hopping
parameter. The optical conductivity displays clearly both the Mott gap
and the incoherent Hubbard bands.

In this slave-fermion description, the Mott-gap state is accompanied by
the formation of holon-doublon bound states. Holon-doublon binding provides
the energy scale of the Mott gap in the half-filled Hubbard model. In greater 
detail, the pairing interaction of holons and doublons is a $k$-space 
phenomenon, occurring across the Fermi surface. We comment that both holons 
and doublons are components of the quasiparticles making up the lower Hubbard 
band, as they are of the upper, and thus the paired state should not be 
considered as an exciton; indeed, holon-doublon pairs would remain present 
at finite doping, when the particle-hole symmetry of the half-filled system 
is lost. The presence of these bound states, together with the interactions 
between the charge carriers and the magnons, reflects the fact that the high- 
and low-energy degrees of freedom in the Hubbard model are intrinsically mixed, 
with the spin fluctuations playing an essential role in the construction of 
coherent ``spin-polaron'' features in the lower and upper Hubbard bands.

Our technique is a strong-coupling approach, valid at large values of $U$,
and can be used to obtain the analytical limiting forms of all physical
quantities in this regime. At intermediate $U$ values, it provides accurate
results for the evolution of the charge dynamics as a consequence of the
renormalising effects of spin fluctuations. Only on the approach to the
weak-coupling regime do we find systematic discrepancies in the framework,
which we may characterise by violations of the spectral sum rule. In this 
regime, the single-mode approximation to the spin dynamics of the N\'eel
antiferromagnetic ground state begins to break down, the self-consistent 
Born approximation requires the inclusion of vertex corrections and even 
the decoupling of electron operators into separate spin and charge parts 
may no longer be justified. As noted in section~\ref{Sec1}, a degree of 
confusion exists in the literature over whether the insulating properties 
of the weakly coupled system can be ascribed to magnetic rather than to 
electrostatic interactions, and we conclude from the absence of a transition 
that the two pictures are different sides of the same coin, driven by the 
same fundamental processes. However, the slave-fermion framework adopted 
here is clearly not the appropriate method for investigating the crossover 
from the robustly Mott-insulating regime to the noninteracting limit.

Despite this deficiency far from its regime of validity, we have
demonstrated that the holon-doublon representation provides both a full
qualitative understanding of the underlying physics of the Mott insulator
and a semi-quantitative account of its physical properties across the full
range of intermediate and strong interactions. Our results for the Mott state
have a transparent origin, not obscured by the complexities of a many-body
numerical calculation, and indeed can be used to verify the features and
parameter-dependences found in numerical studies. Although one may argue that 
the half-filled system at zero temperature is already fully characterised, 
the holon-doublon description we have introduced here provides a valuable 
foundation for understanding both the charge dynamics of the half-filled 
Hubbard model at finite temperatures and the evolution of the Mott gap as 
the system is doped away from half-filling.

\section*{Acknowledgments}

We thank R. Yu for valuable discussions.
This work was supported by the National Natural Science Foundation of China
(Grant Nos.~10934008, 10874215, and 11174365) and by the National Basic
Research Program of China (Grant Nos.~2012CB921704 and 2011CB309703).

\appendix
\section{Hubbard Hamiltonian}
\label{ahh}

\subsection{Heisenberg Model}
\label{Heisenberg}

The spin degrees of freedom are assumed to be governed by the Heisenberg
model, which is treated in the linear spin-wave approximation using the
condensation assumption of equation~(\ref{Condensation}). Despite the extreme
quantum nature of the $S = 1/2$ spins, a linear spin-wave treatment has
been shown \cite{Auerbach-1994} to provide a good description of the
ordered phase with only small, quantitative, and well characterised
renormalisation parameters. In momentum space, the Hamiltonian can be
expressed as
\begin{equation}
H_{S} = {\textstyle \frac{1}{4}} Jzb_{0}^{2} \sum_{\mathbf{k}}(\!
\begin{array}[c]{cc}
s_{\mathbf{k}}^{\dag} & s_{-\mathbf{k}}
\end{array}
\!) \left(\!\!
\begin{array}[c]{cc}
1 & \gamma_{\mathbf{k}} \\ \gamma_{\mathbf{k}} & 1
\end{array}
\!\!\right)  \left(\!\!
\begin{array}
[c]{c}
s_{\mathbf{k}} \\ s_{-\mathbf{k}}^{\dag}
\end{array}
\!\! \right) + C,
\end{equation}
where $C$ is a constant. The Bogoliubov transformation
\begin{equation}
\left(\!\!
\begin{array}[c]{c}
s_{\mathbf{k}} \\ s_{-\mathbf{k}}^{\dag}%
\end{array}
\!\!\right) = \left(\!\!
\begin{array}[c]{cc}
u_{\mathbf{k}} & v_{\mathbf{k}} \\
v_{\mathbf{k}} & u_{\mathbf{k}}
\end{array}
\!\!\right)  \left(\!\!
\begin{array}[c]{c}
\alpha_{\mathbf{k}} \\ \alpha_{-\mathbf{k}}^{\dag}
\end{array}
\!\!\right)  \!,
\end{equation}
with
\begin{equation}
u_{\mathbf{k}}^{2} = \frac{1}{2} + \frac{1}{2\omega_{\mathbf{k}}}, \; v_{\mathbf{k}}^{2}
 = - \frac{1}{2} + \frac{1}{2\omega_{\mathbf{k}}}, \; u_{\mathbf{k}} v_{\mathbf{k}} =
 - \frac{\gamma_{\mathbf{k}}}{2\omega_{\mathbf{k}}}, \; \omega_{\mathbf{k}} = \sqrt{1
 - \gamma_{\mathbf{k}}^{2}}
\label{Bogo}
\end{equation}
and
\begin{equation}
\gamma_{\mathbf{k}} = \frac{1}{z} \sum_{\boldsymbol{\delta}} 
e^{i\mathbf{k\cdot}\boldsymbol{\delta}},
\end{equation}
in which $z = 4$ is the coordination number and $\boldsymbol{\delta} = (\pm a,
0)$ and $(0, \pm a)$, yields the form
\begin{equation}
H_{S} = \sum_{\mathbf{k}} \Omega_{\mathbf{k}} (\alpha_{\mathbf{k}}^{\dag}
\alpha_{\mathbf{k}} + {\textstyle \frac{1}{2}}),
\end{equation}
where
\begin{equation}
\Omega_{\mathbf{k}} = {\textstyle \frac{1}{2}} J z b_{0}^{2} \omega_{\mathbf{k}}
\end{equation}
is the magnon dispersion relation. In this approximation, the energy of the
magnon is reduced by the self-consistent condensation parameter $b_{0}^{2}$,
reflecting the renormalisation of the effective Heisenberg coupling constant
caused by the on-site interaction, $U$.

The two types of magnon Green function are defined by
\begin{eqnarray}
D_{1} \left( \mathbf{q}, \tau \right) & = - \left \langle T_{\tau}
\alpha_{\mathbf{q}} \left( \tau \right) \alpha_{\mathbf{q}}^{\dag} \left(
0 \right) \right \rangle \!, \nonumber \\
D_{2} \left( \mathbf{q}, \tau \right) & = - \left \langle T_{\tau}
\alpha_{\mathbf{q}}^{\dag} \left( \tau \right) \alpha_{\mathbf{q}} \left(
0 \right) \right \rangle \!,
\end{eqnarray}
and their Fourier transforms are
\begin{equation}
D_{1} \left( \mathbf{q}, i\omega_{n} \right) = \frac{1}{i\omega_{n}
 - \Omega_{\mathbf{q}}}, \;\; D_{2} \left( \mathbf{q}, i\omega_{n} \right)
 = - \frac{1}{i\omega_{n} + \Omega_{\mathbf{q}}}.
\end{equation}
In figure~\ref{Feynman}, $\mathbf{D}$ is a $2\times2$ matrix whose explicit
form is
\begin{eqnarray}
\mathbf{D}_{11} & = g_{1}^{2} \left( \mathbf{k,q} \right) D_{1} + g_{2}^{2}
\left( \mathbf{k,q} \right) D_{2}, \nonumber \\
\mathbf{D}_{12} & = - g_{1} \left( \mathbf{k,q} \right) g_{2} \left(
\mathbf{k,q} \right) [D_{1} + D_{2}], \nonumber \\
\mathbf{D}_{21} & = - g_{1} \left( \mathbf{k,q} \right) g_{2} \left(
\mathbf{k,q} \right) [D_{1} + D_{2}], \nonumber \\
\mathbf{D}_{22} & = g_{2}^{2} \left( \mathbf{k,q} \right) D_{1} + g_{1}^{2}
\left( \mathbf{k,q} \right) D_{2},
\end{eqnarray}
where
\begin{eqnarray}
g_{1} (\mathbf{k,q}) & = {\textstyle \frac{1}{\sqrt{N}}}
t b_{0} z (u_{q} \gamma_{q-k} + v_{q} \gamma_{k}), \nonumber \\
g_{2} (\mathbf{k,q}) & = {\textstyle \frac{1}{\sqrt{N}}}
t b_{0} z (u_{q} \gamma_{k} + v_{q} \gamma_{k-q}).
\label{gfunc}
\end{eqnarray}

\subsection{Hubbard model}
\label{Hubbard}

Within the same condensation approximation, the full Hubbard Hamiltonian
(\ref{HTU}) is given by
\begin{eqnarray}
H& ={\textstyle\frac{1}{2}}U\sum_{\mathbf{k}}(d_{\mathbf{k}}^{\dagger }d_{%
\mathbf{k}}+e_{\mathbf{k}}^{\dagger }e_{\mathbf{k}})   \nonumber\\
& \quad +4tb_{0}^{2}\sum_{\mathbf{k}}(d_{\mathbf{Q}-\mathbf{k}}^{\dagger }e_{%
\mathbf{k}}^{\dagger }\gamma _{\mathbf{k}}+e_{\mathbf{k}}d_{\mathbf{Q}-%
\mathbf{k}}\gamma _{\mathbf{k}})   \nonumber\\
& \quad -\frac{tb_{0}z}{\sqrt{N}}\sum_{\mathbf{k},\mathbf{q}}d_{\mathbf{k}%
}^{\dagger }d_{\mathbf{k-q}}(\gamma _{\mathbf{k}}s_{-\mathbf{q}}^{\dagger
}+\gamma _{\mathbf{k}-\mathbf{q}}s_{\mathbf{q}})   \nonumber\\
& \quad +\frac{tb_{0}z}{\sqrt{N}}\sum_{\mathbf{k},\mathbf{q}}e_{\mathbf{k}%
}^{\dagger }e_{\mathbf{k}-\mathbf{q}}(\gamma _{\mathbf{k}}s_{-\mathbf{q}%
}^{\dagger }+\gamma _{\mathbf{k}-\mathbf{q}}s_{\mathbf{q}}).
\label{Momentum}
\end{eqnarray}
The first line expresses a mass term for the holons and doublons, whose
pairing interaction, in the second line, gives them a dispersive kinetic
term in the unperturbed charge Green function. The factor of $b_{0}^{2}$
contains the renormalisation of the pairing strength due to spin fluctuations.
The third and fourth lines of equation~(\ref{Momentum}) express the fact that 
the hopping of holons and doublons is coupled with the emission and absorption
of magnons, and we treat this three-body scattering interaction by the
SCBA.

In terms of the Nambu spinor, $\psi_{\mathbf{k}}^{\dagger} = (\!\!
\begin{array}[c]{cc}
d_{-\mathbf{k} + \mathbf{Q}}^{\dagger} & e_{\mathbf{k}}
\end{array}
\!\!)$, the Hamiltonian takes the matrix form
\begin{equation}
H = H_{C} + H_{CS},
\label{ehccs}
\end{equation}
in which
\begin{equation}
H_{C} = \sum_{\mathbf{k}} \psi_{\mathbf{k}}^{\dagger} \tilde{\varepsilon}_{\mathbf{k}}
\psi_{\mathbf{k}}
\label{ehc}
\end{equation}
describes the charge dynamics, contained in the first two lines of
equation~(\ref{Momentum}), while the interaction between the charge and
spin degrees of freedom (latter two lines) is given by
\begin{equation}
H_{CS} = \sum_{\mathbf{k},\mathbf{q}} \psi_{\mathbf{k}}^{\dagger} M (\mathbf{k},
\mathbf{q}) \psi_{\mathbf{k}-\mathbf{q}};
\label{ehcs}
\end{equation}
in equations~(\ref{ehc}) and (\ref{ehcs}) we have defined
\begin{eqnarray}
\tilde{\varepsilon}_{\mathbf{k}} = \left(\!\!
\begin{array}[c]{cc}
U/2 & 4t b_{0}^{2} \gamma_{\mathbf{k}} \\ 4t b_{0}^{2} \gamma_{\mathbf{k}} & - U/2
\end{array}
\!\!\right) \!\!, \nonumber \\ M (\mathbf{k},\mathbf{q}) = \left(\!\!
\begin{array}[c]{cc}
M_{11} (\mathbf{k},\mathbf{q})  & 0 \\ 0 & M_{22} (\mathbf{k},\mathbf{q})
\end{array}
\!\!\right) \!\!, \nonumber \\
M_{11} (\mathbf{k},\mathbf{q}) = g_{2} (\mathbf{k,q}) \alpha_{\mathbf{q}}^{\dag}
 + g_{1} (\mathbf{k,q}) \alpha_{-\mathbf{q}}, \nonumber \\
M_{22} (\mathbf{k},\mathbf{q}) = - g_{1} (\mathbf{k,q}) \alpha_{\mathbf{q}}^{\dag}
 - g_{2} (\mathbf{k,q}) \alpha_{-\mathbf{q}}. \nonumber
\end{eqnarray}
The unperturbed charge Green function is given by
\begin{eqnarray}
\mathbf{F}^{(0)} (\mathbf{k},i\omega_{n}) & = [i\omega_{n} -
\tilde{\varepsilon}_{\mathbf{k}}]^{-1} \\
& = \frac{1}{(i\omega_{n} - E_{\mathbf{k}}) (i\omega_{n} + E_{\mathbf{k}})}
\left(\!\!
\begin{array}[c]{cc}
i\omega_{n} + U/2 & 4t b_{0}^{2} \gamma_{\mathbf{k}} \\
4t b_{0}^{2} \gamma_{\mathbf{k}} & i\omega_{n} - U/2
\end{array} \!\!\right) \!\!, \nonumber
\label{ZeroHamil}
\end{eqnarray}
where
\begin{equation}
E_{\mathbf{k}} = \sqrt{{\textstyle \frac{1}{4}} U^2 + (4t b_{0}^{2}
\gamma_{\mathbf{k}})^{2}}
\label{edisp}
\end{equation}
is the quasiparticle dispersion relation for the charge degrees of freedom.

The self-consistent Dyson equation for the full Green function in the presence
of spin-fluctuation interactions is calculated in the SCBA to deduce the
Green function,
\begin{equation}
\mathbf{F} (\mathbf{k},i\omega_{n}) = \frac{1}{i\omega_{n} -
\tilde{\varepsilon}_{\mathbf{k}} - \mathbf{\Sigma} (\mathbf{k},i\omega_{n})},
\label{ecgf}
\end{equation}
of equation~(\ref{egf}) with the self-energy given by
\begin{eqnarray}
\mathbf{\Sigma} (\mathbf{k},i\omega_{n}) 
& = \sum_{\mathbf{q}} \int_{-\infty}^{\infty} d\varepsilon \frac{1 - f(\varepsilon)}
{i\omega_{n} - \varepsilon - \Omega_{\mathbf{q}}} \mathbf{g} (\mathbf{k,q})
\mathbf{A}_{1} (\mathbf{k} - \mathbf{q}, \varepsilon) \mathbf{g} (\mathbf{k,q})
\nonumber \\
& + \sum_{\mathbf{q}} \int_{-\infty}^{\infty} d\varepsilon \frac{f(\varepsilon)}
{i\omega_{n} - \varepsilon + \Omega_{\mathbf{q}}} \mathbf{g} (\mathbf{k,q})
\mathbf{A}_{2} (\mathbf{k} - \mathbf{q}, \varepsilon) \mathbf{g} (\mathbf{k,q}),
\label{SelfE}
\end{eqnarray}
in which
\begin{equation}
\mathbf{g} (\mathbf{k,q}) = \left[
\begin{array}[c]{cc}
g_{1} (\mathbf{k,q}) & 0 \\ 0 & - g_{2} (\mathbf{k,q})
\end{array}
\right] \!\! ,
\end{equation}
and
\begin{eqnarray}
\mathbf{A}_{1} (\mathbf{k},\varepsilon) & = & \left[
\begin{array}[c]{cc}
A_{11} (\mathbf{k},\varepsilon) & A_{12} (\mathbf{k},\varepsilon) \\
A_{21} (\mathbf{k},\varepsilon) & A_{22} (\mathbf{k},\varepsilon)
\end{array}
\right] \!\! , \\
\mathbf{A}_{2} (\mathbf{k},\varepsilon) & = & \left[
\begin{array}[c]{cc}
A_{22} (\mathbf{k},\varepsilon) & A_{12} (\mathbf{k},\varepsilon) \\
A_{21} (\mathbf{k},\varepsilon) & A_{11} (\mathbf{k},\varepsilon)
\end{array}
\right] \!\! ,
\end{eqnarray}
with
\begin{equation}
A_{\alpha\beta} (\mathbf{k},\varepsilon) = - \frac{1}{\pi} \rm{I}\rm{m}
F_{\alpha\beta}^{R} (\mathbf{k},\varepsilon + i \eta)
\end{equation}
the spectral function corresponding to the retarded Green function and
$f(\varepsilon)$ the Fermi distribution function.

\section{Large-$U$ Limit}
\label{largeulimit}

Because the slave-fermion treatment of the Hubbard model is exact in the
limit of strong on-site Coulomb repulsion, the SCBA affords valuable insight
both into the limiting properties and into the effects of charge and spin
fluctuations away from the limiting regime. In this appendix we review the
magnetic order, Green function and approximate spectral properties at large
$U$.

\subsection{Bose Condensation and Magnetic Order}
\label{lum}

In the slave-fermion decomposition, the self-consistent condition has the
form [equation~(\ref{Consist})]
\begin{equation}
\frac{1}{N} \sum_{\mathbf{k}} \left[ \langle d_{\mathbf{k}}^{\dagger } d_{\mathbf{k}}
\rangle + \langle e_{\mathbf{k}}^{\dagger } e_{\mathbf{k}} \rangle + v_{\mathbf{k}}^{2}
\right] + b_{0}^{2} = 1,
\end{equation}
where the coefficient $v_{\mathbf{k}}^{2}$ of the Bogoliubov transformation is
defined in equation~(\ref{Bogo}). At half-filling and in the large-$U$ limit,
$\sum_{\mathbf{k}} \langle d_{\mathbf{k}}^{\dagger } d_{\mathbf{k}} \rangle =
\sum_{\mathbf{k}} \langle e_{\mathbf{k}}^{\dagger } e_{\mathbf{k}} \rangle = 0$, leaving
\begin{equation}
b_{0}^{2} = 1 - \frac{1}{N} \sum_{\mathbf{k}} v_{\mathbf{k}}^{2} \simeq 0.803
\end{equation}
for the nearest-neighbour kinetic term.

In the Schwinger-boson representation, the staggered magnetisation of the
magnetically ordered phase is related to this condensed fraction by
\begin{eqnarray}
m_{s} & = & \frac{1}{N} \sum_{i} \langle S_{i}^{z} \rangle = \frac{1}{N}
\sum_{i} \frac{1}{2} (\langle s_{i,\uparrow }^{\dag } s_{i,\uparrow} \rangle
 - \langle s_{i,\downarrow }^{\dag } s_{i,\downarrow} \rangle) \nonumber \\
& = & \frac{1}{2} b_{0}^{2} - \frac{1}{2N} \sum_{\mathbf{k}} v_{\mathbf{k}}^{2} \simeq
0.303.
\end{eqnarray}
One observes that this treatment of the $S = 1/2$ system yields a remarkably
accurate reproduction of the known ordered moment, $m_s \simeq 0.306$,
computed by unbiased numerical methods \cite{Sandvik-1997}. Our calculations
are performed on a square lattice of size $48 \times 48$ sites with periodic
boundary conditions, and due to this finite-size effect give the approximate
values
\begin{equation}
b_{0}^{2} \simeq 0.821, \;\;\;\; m_{s} \simeq 0.321.
\end{equation}

\subsection{Green Function and Double Occupancy}
\label{lud}

In the strong-coupling limit, to lowest order one may consider only the
charge part of the Hamiltonian, $H_C$ in equation~(\ref{ehccs}), and neglect 
the terms $H_{CS}$ describing the interaction of the charge degrees of freedom
with the spin fluctuations. In this approximation, the doublon Green function
\begin{equation}
\mathbf{F}_{11}^{(0)} (\mathbf{k},\tau) = - \left\langle T_{\tau } \left[
d_{-\mathbf{k}+\mathbf{Q}} (\tau) d_{-\mathbf{k}+\mathbf{Q}}^{\dagger } (0) \right]
\right\rangle \!,
\end{equation}
may be expressed as
\begin{equation}
\mathbf{F}_{11}^{(0)} (\mathbf{k},i\omega_{n}) = \frac{i\omega _{n} + U/2}
{(i\omega_{n} - E_{\mathbf{k}}) (i\omega_{n} + E_{\mathbf{k}})},
\label{edgf}
\end{equation}
with $E_{\mathbf{k}}$ given in equation~(\ref{edisp}).

By rewriting equation~(\ref{edgf}) in the form
\begin{eqnarray}
\mathbf{F}_{11}^{(0)} (\mathbf{k},\omega +i\delta) & = & \frac{E_{\mathbf{k}}
 + U/2}{2E_{\mathbf{k}}} \frac{1}{\omega + i\delta - E_{\mathbf{k}}} + \, 
\frac{E_{\mathbf{k}} - U/2}{2E_{\mathbf{k}}} \frac{1}{\omega + i\delta
 + E_{\mathbf{k}}},
\end{eqnarray}
one obtains the spectral function
\begin{eqnarray}
\mathbf{A}_{11}^{(0)} (\mathbf{k},\omega) & = & \frac{E_{\mathbf{k}} + U/2}
{2E_{\mathbf{k}}} \delta (\omega - E_{\mathbf{k}}) + \, \frac{E_{\mathbf{k}} - U/2}
{2E_{\mathbf{k}}} \delta (\omega + E_{\mathbf{k}})
\label{edsf}
\end{eqnarray}
and thus the zero-temperature doublon occupation function
\begin{equation}
\langle d_{-\mathbf{k}+\mathbf{Q}}^{\dagger} d_{-\mathbf{k}+\mathbf{Q}} \rangle = 
\int_{-\infty }^{\infty } \!\! d\omega f (\omega) A_{11}^{(0)} (\mathbf{k},\omega)
 = \frac{E_{\mathbf{k}} - \frac{U}{2}}{2E_{\mathbf{k}}}.
\end{equation}
By Taylor expansion of this expression,
\begin{eqnarray}
\langle d_{-\mathbf{k}+\mathbf{Q}}^{\dagger} d_{-\mathbf{k}+\mathbf{Q}} \rangle
& \simeq & \frac{1}{U} \left\{ \frac{U}{2} \left[ 1 + \frac{1}{2}
\frac{4}{U^{2}} (4 t b_{0}^{2} \gamma_{\mathbf{k}})^{2} \right] - \frac{U}{2}
\right\} \nonumber \\ & \simeq & (4 t b_{0}^{2} \gamma_{\mathbf{k}})^{2}
\frac{1}{U^{2}}
\end{eqnarray}
and the net doublon occupation is
\begin{equation}
D \simeq \frac{1}{U^{2}} \frac{1}{N} \sum_{\mathbf{k}} (4 t b_{0}^{2}
\gamma_{\mathbf{k}})^{2} \simeq \frac{2.70}{U^{2}},
\end{equation}
showing asymptotic $1/U^2$ scaling in the large-$U$ limit.

\subsection{Mott Gap}
\label{lug}

In the limit of large $U$, one may assume that the strongest contributions
to the renormalisation of the Mott gap are obtained at first order in the
SCBA, with higher-order contributions being heavily suppressed. In this case
it is safe to assume that the off-diagonal Green-function components,
$\mathbf{F}_{12}$ and $\mathbf{F}_{21}$ in equation~(\ref{ecgf}), may be neglected
in computing the self-energy corrections determining the Mott gap. From the
doublon spectral function of equation~(\ref{edsf}) and its holon counterpart
\begin{eqnarray}
\mathbf{A}_{22}^{(0)} (\mathbf{k},\omega) & = & \frac{E_{\mathbf{k}} - U/2}
{2E_{\mathbf{k}}} \delta (\omega - E_{\mathbf{k}}) + \, \frac{E_{\mathbf{k}} + U/2}
{2E_{\mathbf{k}}} \delta (\omega + E_{\mathbf{k}}),
\label{ehsf}
\end{eqnarray}
the self-energy components [equation~(\ref{SelfE})] calculated in the SCBA are
\begin{eqnarray}
\mathbf{\Sigma }_{11} (\mathbf{k},i\omega_{n}) & = & \sum_{\mathbf{q}} g_{1}^{2}
(\mathbf{k,q}) \int_{0}^{\infty } d\varepsilon \frac{A_{11}^{(0)}(\mathbf{k}
 - \mathbf{q},\varepsilon )}{i\omega_{n}-\Omega_{\mathbf{q}} - \varepsilon}
[1 - f(\varepsilon )] \nonumber \\
& & + \sum_{\mathbf{q}} g_{2}^{2}(\mathbf{k,q}) \int_{-\infty }^{0} d\varepsilon
\frac{A_{11}^{(0)}(\mathbf{k} - \mathbf{q},\varepsilon )}{i\omega _{n} + 
\Omega_{\mathbf{q}}-\varepsilon} f(\varepsilon) \nonumber \\
& = & \sum_{\mathbf{q}} \frac{E_{\mathbf{k}-\mathbf{q}} + U/2}{2E_{\mathbf{k}
 - \mathbf{q}}} g_{1}^{2} (\mathbf{k,q}) \frac{1}{i\omega_{n} - \Omega_{q}
 - E_{\mathbf{k}-\mathbf{q}}}  \nonumber \\
& & + \sum_{\mathbf{q}} \frac{E_{\mathbf{k}-\mathbf{q}} - U/2}{2E_{\mathbf{k} 
 - \mathbf{q}}} g_{2}^{2} (\mathbf{k,q}) \frac{1}{i\omega _{n} + \Omega_{q}
 + E_{\mathbf{k}-\mathbf{q}}}  \nonumber \\
& \approx & \sum_{\mathbf{q}} g_{1}^{2} (\mathbf{k,q}) \frac{1}{i\omega _{n} - U/2}
\; = \; \frac{a_{\mathbf{k}}}{i\omega _{n}-U/2},  
\label{DSESCBA}
\end{eqnarray}
and
\begin{eqnarray}
\mathbf{\Sigma }_{22} (\mathbf{k},i\omega_{n}) & = & \sum_{\mathbf{q}} g_{2}^{2}
(\mathbf{k,q}) \int_{0}^{\infty } d\varepsilon \frac{A_{22}^{(0)} (\mathbf{k}
 - \mathbf{q},\varepsilon )}{i\omega_{n} - \Omega_{q} - \varepsilon}
[1 - f(\varepsilon )] \nonumber \\
& & + \sum_{\mathbf{q}} g_{1}^{2}(\mathbf{k,q}) \int_{-\infty }^{0} d\varepsilon
\frac{A_{22}^{(0)} (\mathbf{k} - \mathbf{q},\varepsilon)}{i\omega_{n} + 
\Omega_{q} - \varepsilon } f(\varepsilon )  \nonumber \\
& = & \sum_{\mathbf{q}} \frac{E_{\mathbf{k} - \mathbf{q}} - U/2}{2E_{\mathbf{k}
 - \mathbf{q}}} g_{2}^{2} (\mathbf{k,q}) \frac{1}{i\omega_{n} - \Omega_{q}
 - E_{\mathbf{k} - \mathbf{q}}}  \nonumber \\
& & + \sum_{\mathbf{q}} \frac{E_{\mathbf{k} - \mathbf{q}} + U/2}{2E_{\mathbf{k}
 - \mathbf{q}}} g_{1}^{2} (\mathbf{k,q}) \frac{1}{i\omega _{n} + \Omega_{q}
 + E_{\mathbf{k} - \mathbf{q}}} \nonumber \\
& \approx & \sum_{\mathbf{q}} g_{1}^{2}(\mathbf{k,q}) \frac{1}{i\omega_{n} + U/2}
\; = \; \frac{a_{\mathbf{k}}}{i\omega _{n}+U/2},  \label{HSESCBA}
\end{eqnarray}
where
\begin{equation}
a_{\mathbf{k}} = \sum_{\mathbf{q}} g_1^{2} (\mathbf{k,q}).
\label{eag}
\end{equation}

The SCBA Green function [equations~(\ref{ZeroHamil}) and (\ref{ecgf})] is then
\begin{equation}
\mathbf{F} (\mathbf{k},i\omega_{n}) = \left[
\begin{array}{cc}
i\omega_{n} - U/2 - \frac{a_{\mathbf{k}}}{i\omega _{n} - U/2} & - 4 t b_{0}^{2} 
\gamma_{\mathbf{k}} \\ - 4 t b_{0}^{2} \gamma_{\mathbf{k}} & i\omega_{n} + U/2
 - \frac{a_{\mathbf{k}}}{i\omega_{n} + U/2} \! \end{array} \right]^{\!-1} \!, 
\nonumber
\label{ZeroHamil2}
\end{equation}
whose doublon and holon components are
\begin{eqnarray}
\fl \mathbf{F}_{11} (\mathbf{k},i\omega_{n}) = \label{DSCBA} \frac{i\omega_{n}
 + U/2 - \frac{a_{\mathbf{k}}}{i\omega_{n} + U/2}}{\left[ i\omega_{n} - U/2
 - \frac{a_{\mathbf{k}}}{i\omega_{n} - U/2}\right] \! \left[i\omega_{n} + U/2
 - \frac{a_{\mathbf{k}}}{i\omega_{n} + U/2}\right] - (4 t b_{0}^{2} 
\gamma_{\mathbf{k}})^{2}},  \\
\fl \mathbf{F}_{22} (\mathbf{k},i\omega_{n}) = \label{HSCBA} \frac{i\omega_{n}
 - U/2 - \frac{a_{\mathbf{k}}}{i\omega_{n} + U/2}}{\left[ i\omega_{n} - U/2
 - \frac{a_{\mathbf{k}}}{i\omega_{n} - U/2}\right] \! \left[i\omega_{n} + U/2
 - \frac{a_{\mathbf{k}}}{i\omega_{n} + U/2}\right] - (4 t b_{0}^{2}
\gamma_{\mathbf{k}})^{2}}.
\end{eqnarray}
These functions both have four poles, which are the same in both cases,
\begin{equation}
E_k^m = \pm \frac12 \sqrt{4 a_{\mathbf{k}} + 2 b_{\mathbf{k}} + U^2 \pm 2
\sqrt{4 a_{\mathbf{k}} b_{\mathbf{k}} + b_{\mathbf{k}}^2 + 4 a_{\mathbf{k}} U^2}},
\label{efpdhgf}
\end{equation}
where we have defined $b_{\mathbf{k}} = (4 t b_0^2 \gamma_{\mathbf{k}})^2$. Thus
the holon and doublon gaps are both given by
\begin{equation}
\Delta_{e,\mathbf{k}} = \Delta_{d,\mathbf{k}} \approx U - 2\sqrt{a_{\mathbf{k}}}
\end{equation}
in the limit of large $U$.

The electron Green function is the convolution of the holon and doublon
Green functions, which include the magnon renormalisation effects, and
hence the Mott gap is given by the holon and doublon gaps. From
figure~\ref{Resolved}(a), it is clear that the minimum value of this gap
corresponds to the high density of states giving the clear peaks in
figure~\ref{Spectral}(b). Thus by setting $\mathbf{k} = (\pi/2,\pi/2)$ and
performing the $\mathbf{q}$-sum over the Brillouin zone within $a_{\mathbf{k}}$
(\ref{eag}) we obtain (main text)
\begin{equation}
\Delta = {\rm min}|_{\mathbf{k}} \left[ U - 2 \sqrt{a_{\mathbf{k}}} \right]
 \; = \; U - 3.77
\end{equation}
in the strong-coupling limit. We conclude that the approach of the Mott gap,
$\Delta/U = 1 - 3.77/U$, to unity due to the suppression of spin fluctuations
in this regime is proportional to $1/U$. Because this derivation considers
only the first order in the self-consistent renormalisation scheme, one may
expect the coefficient to be larger than $3.77$ if higher-order processes
are included.

\section{Optical conductivity}
\label{Opticalconductivity}

The conductivity tensor $\sigma_{\mu\nu} (\mathbf{q},\omega)$ is defined as
the linear response of the charge current density in a solid to the total
electric field,
\begin{equation}
\left\langle \mathcal{\hat{J}}_{\mu} (\mathbf{q},\omega) \right\rangle
 = \sum_{\nu} \sigma_{\mu\nu} (\mathbf{q},\omega) E_{\nu} (\mathbf{q},\omega),
\end{equation}
where $\mu$ and $\nu$ denote Cartesian coordinates and $\left\langle
\mathcal{\dots} \right\rangle$ denotes the equilibrium average; we use
units where $e = 1$, $\hbar = 1$ and $c = 1$. The long-wavelength
$(\mathbf{q} = 0)$ limit of the real part, \textrm{Re}$[\sigma_{\mu\nu}
(\omega)]$, is referred to as the optical conductivity, because the
wavelength of electromagnetic waves far exceeds the characteristic length
scales of condensed matter systems, so here we derive this limit.

For a lattice system, the hopping parameter in the presence of an
electromagnetic field is expressed by the Peierls substitution
\cite{Bergeron-2011} as
\begin{equation}
t_{ij} = t_{ij}^{0} \rm{e}\rm{x}\rm{p} \left[ i \int_{i}^{j} d\mathbf{l}
\cdot \mathbf{A} (\mathbf{r},t) \right] \!\!,
\end{equation}
where $\int_{i}^{j} d \mathbf{l} \cdot \mathbf{A} (\mathbf{r},t)$ is the 
integral of the vector potential along the hopping path. For a spatially 
uniform electric field, the vector potential can be taken as independent 
of position, $\mathbf{A}(\mathbf{r},t) \rightarrow \mathbf{A} (t)$. By 
restricting to the linear-response regime, expanding the phase factors 
to second order in $\mathbf{A} (t)$ and taking the Fourier transform, 
one obtains the current operator in the form
\begin{equation}
\mathcal{\hat{J}}_{\mu} (\mathbf{k},t) = \mathcal{\hat{J}}_{\mu}^{(1)}
(\mathbf{k},t) + \mathcal{\hat{J}}_{\mu}^{(2)} (\mathbf{k},t)  ,
\end{equation}
where
\begin{eqnarray}
& \mathcal{\hat{J}}_{\mu}^{(1)} (\mathbf{q},t) = - \sum_{\mathbf{k},\sigma} 
\frac{\partial\varepsilon_{(2\mathbf{k+q})/2}}{\partial k_{x}} 
c_{\mathbf{k},\sigma}^{\dag} c_{\mathbf{k+q},\sigma}, \\
& \mathcal{\hat{J}}_{\mu}^{(2)} (\mathbf{q},t) = - \sum_{\mathbf{k},\mathbf{p},\sigma} 
\frac{\partial^{2}\varepsilon_{(\mathbf{k+p})/2}}{\partial k_{x}^{2}} 
c_{\mathbf{k},\sigma}^{\dag} c_{\mathbf{p},\sigma} A_{\mathbf{k-p}+\mathbf{q}}^{\mu} (t),
\end{eqnarray}
with $\varepsilon_{\mathbf{k}}$ the electron dispersion relation.

$\mathcal{\hat{J}}_{\mu}^{(1)} (\mathbf{k},t)$ is known as the paramagnetic
current and $\mathcal{\hat{J}}_{\mu}^{(2)} (\mathbf{k},t)$ as the diamagnetic
one. The latter is already linear with respect to the applied field but for
the former the Kubo formula is required. To linear order in the applied field,
the interaction term in the Hamiltonian is given by
\begin{equation}
H^{\prime} (t) = - \frac{1}{i\omega} \sum_{\mathbf{q}} \mathcal{\hat{J}}_{\mu}^{(1)}
(-\mathbf{q}) \mathbf{E}_{\mathbf{q}}^{\mu} e^{-i \omega t}.
\end{equation}
The imaginary-time current-current correlation function is defined as
\begin{equation}
\Pi_{\alpha\beta} (\mathbf{q},\tau) = - \frac{1}{N} \left\langle T_{\tau} \left[
\mathcal{\hat{J}}_{\alpha}^{(1)} (\mathbf{q},\tau) \mathcal{\hat{J}}_{\beta}^{(1)}
(-\mathbf{q},0) \right] \right\rangle
\end{equation}
and its Fourier transform as
\begin{equation}
\Pi_{\alpha\beta} (\mathbf{q},i\omega_{n}) = \int_{0}^{\beta} d\tau e^{i\omega_{n}\tau}
\Pi_{\alpha\beta} (\mathbf{q},\tau),
\end{equation}
both of which are obtained from the charge Green function of 
equation~(\ref{ecgf}). Finally, the conductivity in the $x$ direction 
is given by 
\begin{eqnarray}
& \sigma_{xx} (\mathbf{q},\omega) = \frac{i}{\omega} \left[ \Pi_{xx}^{R} 
(\mathbf{q},\omega) + \frac{1}{N} \sum_{\mathbf{k},\sigma} \frac{\partial^{2} 
\varepsilon_{\mathbf{k}}}{\partial k_{x}^{2}} \langle c_{\mathbf{k},\sigma}^{\dag} 
c_{\mathbf{k},\sigma} \rangle \right] \!\!,
\end{eqnarray}
whence the real part of the optical conductivity in the limit $\mathbf{q} = 0$
is
\begin{equation}
\mathrm{Re} [\sigma_{xx} (\omega)] = - \mathrm{Im} \left[ \frac{1}{\omega}
\Pi_{xx}^{R} (\omega) \right] \!\!.
\end{equation}

\end{document}